\documentclass[fleqn,usenatbib,useAMS]{mnras}
\usepackage{amsmath}
\usepackage{amssymb}
\usepackage{CJK}
\usepackage{threeparttable}
\usepackage{graphicx}
\usepackage{orcidlink}

\def\vel{\mathbf{v}}
\def\vs{\vel_{\rm{s}}}
\def\va{\vel_{\rm{A}}}
\def\Fc{\mathbf{F}_{\rm c}}
\def\pg{P_{\rm g}}
\def\pcr{P_{\rm c}}
\def\ecr{E_{\rm c}}
\def\vm{V_{\rm m}}
\def\kdiff{\kappa_{\rm diff}}
\def\cs{c_{\rm{s}}}

\def\Lg{L_{\rm g}}
\def\Lc{L_{\rm c}}

\begin{document}
\begin{CJK*}{UTF8}{gbsn}

\title[CR Driven Outflow]{The Launching of Cosmic Ray Driven Outflows}
\author[Huang \& Davis]{Xiaoshan Huang (黄小珊)$^{1}$\orcidlink{0000-0003-2868-489X} \thanks{E-mail: xh2pm@virginia.edu}
Shane W. Davis$^{1}$\orcidlink{0000-0001-7488-4468}
\\
$^{1}$Department of Astronomy, University of Virginia, Charlottesville, VA 22904, USA
}
\date{Accepted 2022 January 6. Received 2022 January 6; in original form 2021 May 21}
\maketitle

\pubyear{2021}

\begin{abstract}

Cosmic rays (CRs) are thought to be an important feedback mechanism in star-forming galaxies.  They can provide an important source of pressure support and possibly drive outflows. We perform multidimensional CR-magnetohydrodynamic simulations including transport by streaming and diffusion to investigate wind launching from an initially hydrostatic atmosphere by CRs. We estimate a characteristic Eddington limit on the CR flux for which the CR force exceeds gravity and compare it to simulated systems. Scaling our results to conditions in star-forming galaxies, we find that CRs are likely to contribute to driving outflows for a broad range of star formation environments. We quantify the momentum and energy transfer between CRs and gas, along with the associated mass outflow rates under different assumptions about the relative importance of streaming and diffusion for transport. In simulations with streaming, we observe the growth and saturation of the CR acoustic instability, but the CRs and gas remain well coupled, with CR momentum transferred efficiently to the gas even when this instability is present. Higher CR fluxes transfer more energy to the gas and drive stronger outflows. When streaming is present, most of the transferred energy takes the form of Alfv\'{e}n wave heating of the gas, raising its pressure and internal energy, with a lower fractional contribution to the kinetic energy of the outflow.  We also consider runs with radiative cooling, which modifies gas temperature and pressure profiles but does not seem to have a large impact on the mass outflow for super-Eddington CR fluxes.

\end{abstract}


\section{Introduction}\label{sec:intro}

The general picture that stars are formed from collapsing gas is widely agreed upon. However, it has long been a puzzle that the observed star formation rate (SFR) is usually below what is predicted by the simple assumption of gravitational collapse of gas clouds on dynamical timescales \citep{kennicutt1998global, krumholz2007slow, kennicutt2012star}. Cosmological simulations resolving galaxy formation indicate that the energy and momentum feedback from star formation is a key component for explaining the inefficient conversion from gas to stars \citep{somerville2015physical}. Such feedback can happen via numerous channels. Galactic outflow, which is observed in wide range of galaxies \citep{veilleux2005galactic, spilker2020ubiquitous, veilleux2020cool}, is thought to be an important feedback mechanism for removing gas from galaxies, shaping the circum-galactic/inter-galactic medium, and modifying galaxy evolution. 

The origin of galactic outflow is not completely understood.  There are several promising mechanisms for launching outflows, but their relative importance in different types of galaxies is still a matter of debate \citep{zhang2018review}. Although there is evidence AGN may be relevant in some cases, the outflows correlate with star formation activity \citep{martin2005mapping, veilleux2005galactic}, implying that mechanisms tied to star-formation operate. Hot supernova-driven thermal winds are almost certainly present in some galaxies \citep{chevalierclegg85}.  Entrainment of colder gas in these hot winds \citep{strickland2009supernova, scannapieco2015launching, fielding2018clustered, gronke2020cold}, and radiation pressure on dust \citep{2012ApJ...760..155K, zhang2018dusty, huang2020dusty,kim2021star} are promising mechanisms for explaining the molecular and weakly ionised gas that is observed. We refer to the latter mechanism as non-thermal feedback since relativistic particles (photons) are doing the driving rather than hot gas.

Another non-thermal feedback candidate is cosmic ray (CR) pressure. Being roughly in energy equipartition in our local universe \citep{boulares1990galactic,ferriere2001interstellar}, CRs are thought to be an attractive source to drive outflows in Milky Way-like systems \citep{ipavich1975galactic, breitschwerdt1991galactic, breitschwerdt1993galactic, everett2008milky}, as well as in star-bursting systems \citep{socrates2008eddington}. Meanwhile, numerical works also suggest that CR can drive outflow with significant mass-loading and effectively regulate the star formation rate \citep{jubelgas2008cosmic,uhlig2012galactic,booth2013simulations,hanasz2013cosmic,salem2014cosmic}. 

Recent semi-analytical and numerical simulations further elaborate CR's feedback role in terms of driving outflow, modifying galaxy evolution and shaping circum-galactic environment at different scales. For example, cosmological and galaxy formation simulations suggest that when compared to simulations only including thermal feedback, adding CRs changes the star formation history, especially in massive galaxies \citep{ruszkowski2017global, butsky2018role, hopkins2020but}. CRs usually enhance outflow, constrain wind morphology and thicken the galaxy disk \citep{bustard2020cosmic, buck2020effects}. Interestingly, both analytical work and numerical studies suggest that the CR driven wind is likely distinct from a typical momentum or energy driven wind, leaving a unique wind profile \citep{salem2014cosmic,mao2018galactic,girichidis2018cooler}. In the circum-galactic medium (CGM) or intra-galactic medium (IGM), CRs are an important pressure source that supports gas in addition to thermal pressure and potentially allows a cooler halo \citep{butsky2018role, buck2020effects}. The collisionless and collisional loss of CR energy is also an important energy component in the CGM \citep{bustard2020cosmic,crocker2021cosmic}.

The cosmological and galaxy evolution simulations including CRs provide straightforward comparisons with observation, showing that CRs are an important feedback mechanism. In addition, high resolution local simulations with resolved multi-phase gas structure are another channel to diagnose CR physics and its interaction with radiative cooling, conduction, and magnetic field evolution. For example, CRs are able to drive an initially cold cloud that embedded in hot background to roughly the observed outflow velocities without complete disruption by local dynamical instability \citep{2019MNRAS.489..205W, bruggen2020launching}. In the CGM, CRs modify the onset of thermal instability and subsequent multi-phase gas formation \citet{butsky2020impact,kempski2020thermal}.

To connect the cloud-wind interaction scale and the galaxy formation scale, in this work, we investigate the CR's role in launching an outflow from dense atmosphere against gravity near the disk with high resolution simulations. Similar studies of radiative feedback found that the wind-launching process could be impacted by the presence of dynamical instabilities \citep{2012ApJ...760..155K, 2014ApJ...796..107D}. Resolving local dynamics and potential instability \citep{1994ApJ...431..689B, 2018ApJ...860...97H} is an essential step for bridging the gap between well-resolved simulations of individual cloud evolution and incorporating CR to galaxy evolution framework employing sub-grid models.

Despite the importance of CR feedback, our knowledge about the CR transfer mechanism is limited by the lack of direct extra-galactic observation. The canonical treatment of CR transfer usually includes two processes: streaming and diffusion \citep{1971ApJ...170..265S}. The latter process models CR transfer as classical diffusive process with effective diffusivity. Transport by streaming means that CRs stream along the magnetic field and excite waves, with subsequent wave damping enabling energy and momentum exchange with gas. The streaming velocity is usually determined by the balance between wave growth and wave damping, as well as local ionisation condition, ranging from Alfv\'{e}nic to super-Alfv\'{e}nic streaming \citep{1971ApJ...170..265S, holguin2019role, bai2019magnetohydrodynamic, bustard2020cosmicraytransport,zweibel2020role,bambic2021cosmic}. Extensive studies show that the transfer mechanism has significant impact on the disk and halo morphology, the outflow properties and the energetics of CGM and IGM \citep{wiener2013cosmic,ruszkowski2017global,wiener2017cosmic,buck2020effects,ramzan2020outflows}. 

This work aims to study the efficiency of CR driving from a disk atmosphere as a function of a prescribed CR flux, under different assumptions about the transport mechanism.We adopt the two-moment scheme proposed by \citet{2018ApJ...854....5J}, which solves CR transfer equations including both streaming and diffusion self-consistently. We perform a series of adiabatic simulations, primarily with Alfv\'{e}nic streaming and isotropic diffusion to explore how relative importance of streaming and diffusion changes CR feedback.  We also briefly consider the role of radiative cooling on the launching CR winds. A detailed study of radiative cooling in CR driven multi-phase outflow and its impact on cold gas survivability will be presented in a companion paper.

The plan of this paper is as follows: in Section~\ref{sec:numerical_method} we describe the equations solved, simulation set-up and our prescription for a CR Eddington limit. The simulation results are reported in Section~\ref{sec:results}. We discuss the primary implication from our results and relevance to star-forming galaxies in Section~\ref{sec:discussion}, and summarise the main conclusions in Section~\ref{sec:conclusion}.

\section{Numerical Method}\label{sec:numerical_method}
\subsection{Cosmic Ray Transport Equations}
The equations we solve are ideal magneto hydrodynamics (MHD) equations and cosmic rays (CR) transportation equations, which is based on the two-moment scheme proposed by \citet{2018ApJ...854....5J}:
\begin{subequations}
\begin{align}\label{eq:CRMHD}
\frac{\partial\rho}{\partial t}+\nabla\cdot(\rho\vel)=0,&\\
\frac{\partial(\rho\vel)}{\partial t}+\nabla\cdot(\rho\vel\vel-\textbf{BB}+\textsf{P*})=&\nonumber\\
\sigma_{\rm c}\cdot[\Fc-\vel\cdot(E_{\rm c}\textsf{I}+\textsf{P}_{\rm c})]+\rho g&,\label{eq:gasmomentum}\\
\frac{\partial E}{\partial t}+\nabla\cdot[(E+P^{*})\vel-\textbf{B}(\textbf{B}\cdot\vel)]=&\nonumber\\
(\vel+\vs)\cdot\sigma_{c}\cdot[\Fc-\vel\cdot(E_{\rm c}\textsf{I}+\textsf{P}_{\rm c})]+Q,\label{eq:gasenergy}\\
\frac{\partial\textbf{B}}{\partial t}-\nabla\times(\vel\times\textbf{B})=0,&\\
\frac{\partial E_{\rm c}}{\partial t}+\nabla\cdot\Fc=&\nonumber\\
-(\vel+\vs)\cdot\sigma_{c}\cdot[\Fc-\vel\cdot(E_{\rm c}\textsf{I}+\textsf{P}_{\rm c})]\label{eq:crenergy}\\
\frac{1}{\vm^{2}}\frac{\partial\Fc}{\partial t}+\nabla\cdot\textsf{P}_{\rm c}=&\nonumber\\
-\sigma_{c}\cdot[\Fc-\vel\cdot(E_{\rm c}\textsf{I}+\textsf{P}_{\rm c})]&.\label{eq:crmomentum}
\end{align}
\end{subequations}

Here $\rho$, $\vel$, $E$ are fluid density, velocity, total energy, $\textbf{B}$ is magnetic field strength. $P^{*}$ is the sum of gas pressure and magnetic pressure with permeability $\mu=1$. $Q$ is external energy source term, and set to be 0 in adiabatic simulations. 

Equation~\ref{eq:crenergy} and Equation~\ref{eq:crmomentum} are the CR momentum and energy equations. They are the first and second moment integration of the CR advection-diffusion equation, which was derived by \citet{1971ApJ...170..265S}.  $\vs=-\textbf{sgn}(\textbf{B}\cdot\nabla P_{\rm c})\vel_{A}$ is the CR streaming velocity. It has the magnitude of Alfv\'{e}n velocity and directs down the CR pressure gradient. This description of CR transportation is based on the ``self-confinement'' picture. In this framework, when CRs stream through plasma, they excite Alfv\'{e}n waves. When the streaming velocity exceeds the Alfv\'{e}n velocity, the Alfv\'{e}n wave is amplified and eventually saturates. As the Alfv\'{e}n wave grows and perturbs the magnetic field, the irregularities in magnetic field will scatter the CRs and change their pitch angles, reducing the CR streaming velocity to near the Alfv\'{e}n speed. So the CRs stream at local Alfv\'{e}n speed and down the CR pressure gradient. The wave damping enables the momentum and energy exchange between CRs and gas.

$E_{\rm c}$ and $\Fc$ are the cosmic ray energy density and flux. Equation~\ref{eq:crenergy} is widely adopted in CR-MHD simulations. Compared to the one-moment scheme, the two-moment scheme solves the time-dependent Equation~\ref{eq:crmomentum} to update $\Fc$, instead of adopting steady-state CR flux description. Solving separate CR energy and momentum equations makes the code less diffusive because it better handles streaming velocity where the CR pressure gradient is flat. $\vm$ is the maximum CR propagation velocity, it is constant through the whole domain. This is analogous to adopting the reduced speed of light approximation in an explicit two-moment method to solve radiation transfer equation \citep{2013ApJS..206...21S}, $\vm\ll c$ replaces the speed of light $c$ in Equation~\ref{eq:crmomentum} to relax the timestep. As long as $\vm$ is significantly larger than the maximum flow or Alfv\'{e}n velocity, the impact on the dynamics is limited \citep{2018ApJ...854....5J}. 

$\sigma_{c}$ quantifies the CR-gas interaction:
\begin{equation}\label{eq:sigma}
    \sigma_{\rm c}^{-1}=\sigma_{\rm c}^{'-1}+\frac{\textbf{B}}{|\textbf{B}\cdot(\nabla\cdot\textsf{P}_{\rm c})|}\vel_{\rm A}\cdot(E_{\rm c}\textsf{I}+\textsf{P}_{\rm c}),
\end{equation}
where $\sigma_{c}'^{-1}$ is the classical CR diffusion coefficient $\kdiff$. The second terms is an effective streaming coefficient $\sigma_{\rm str}^{-1}$. When $d\Fc/dt$ terms disappears, Equation~\ref{eq:crenergy} reduces to the classic CR energy transport equation \citep{1991A&A...245...79B}. Because the source terms for cosmic rays (Equation~\ref{eq:crenergy},\ref{eq:crmomentum}) and the fluid (Equation~\ref{eq:gasmomentum},\ref{eq:gasenergy}) are the same except for the opposite signs, the conserved variables are the total energy $E+E_{\rm c}$ and momentum $\rho\vel+\Fc/\vm^{2}$. 

\subsection{Eddington flux for Cosmic Rays}\label{subsec:result_eddingtonflux}

A primary goal of this study is to examine the impact of the CR transport mechanism (specifically streaming versus diffusion) on the efficiency of wind launching and acceleration. But, the launching is also sensitive to the flux of CRs through the atmosphere, so we begin by estimating what magnitude of CR flux we expect to be necessary to drive winds in the limit that either streaming or diffusion dominates.  Motivated by \citet{socrates2008eddington}, we formulate this in terms of a CR Eddington flux, where the acceleration due to CR forces just balance the local gravitational acceleration $g$, which we take to be constant. For a given flux $F_{\rm c,x}$ at the base of the atmosphere, we define the CR Eddington flux $F_{\rm Edd}$ as the flux that satisfies $-\partial \pcr/\partial x= -\rho g$, where $\pcr$ is the CR pressure. We assume that the CRs are nearly isotropic so that $\ecr=3\pcr$. 

First we derive an expression for $F_{\rm edd}$ in the CR streaming limit. When streaming dominates the CR transport, $v_{\rm s}=v_{\rm A}$ if $\ecr$ is monotonic. Because $\vm$ is usually significantly larger than any velocity in the simulation, we further assume that the time-dependent term in Equation~(\ref{eq:crmomentum}) is negligible, so that $F_{\rm c,x} \approx 4(v_{\rm s}+v_x)\pcr$, where $v_x$ is the gas velocity along $x$ direction. In other words, the CR flux is the sum of  advective flux and streaming flux. In the launching region of the flow, $v_x$ is small and $v_{\rm A}\gg v_x$, so we further assume that $F_{\rm c,x} \approx 4 v_{\rm A}\pcr$. The CR energy Equation~(\ref{eq:crenergy}) in 1D geometry becomes:
\begin{equation}
    \frac{\partial \ecr}{\partial t}+3v_{\rm A}\frac{\partial\pcr}{\partial x}+4\pcr\frac{\partial v_{\rm A}}{\partial x} = 0,
\end{equation}
In steady state, the time-dependent term vanishes, resulting $\partial(\pcr v_{\rm A}^{4/3})/\partial x=0$.  

With these assumptions, we can approximate 
\begin{equation}
  -\frac{\partial \pcr}{\partial x} \simeq \frac{4 \pcr}{3 v_{\rm A}} \frac{\partial v_{\rm A}}{\partial x}=
   -\frac{2 \pcr}{3 \rho} \frac{\partial \rho}{\partial x},
\end{equation}
where the second equality follows from assuming $\mathbf{B}$ is constant.  Defining $H_{\rho}\equiv (-\partial \rm ln\rho/\partial x)^{-1}$, and setting $\partial \pcr/\partial x= \rho g$, we can solve for
\begin{equation}\label{eq:feddstreaming}
 F_{\rm Edd,str}= 6 v_{\rm A} g \rho H_\rho,
\end{equation}
which we define as the Eddington flux for CRs in the streaming limit.  Note that unlike the expression for radiation with electron scattering opacity, this expression is not constant, depending on the background density and density gradient.  This means that for a given flux, the acceleration of the flow against gravity will generally be most effective at lower densities.  In contrast, to radiation dominated atmospheres, we might expect that streaming CRs will more generically drive outflows, but possibly only in superficially low density regions of the atmosphere.  This may lead to effective acceleration, but only for a relatively small amount of gas.

The diffusion limit is simpler, with $F_{\rm c,x}\approx-\kdiff(\partial\pcr/\partial x)$, where $\kdiff$ is a characteristic CR diffusivity. So the Eddington flux in the diffusion limit is
\begin{equation}\label{eq:fedddiff}
    F_{\rm Edd,diff}\equiv -\kdiff\frac{\partial\pcr}{\partial x}=\kdiff \rho g.
\end{equation}
For $\kdiff$ approximately constant, this is somewhat more like the radiative Eddington limit, but with an opacity that is inversely proportional to density, so the result again suggest that lower density gas will be more readily accelerated.

\subsection{Simulation Setup}

\subsubsection{Units and Scaling}\label{subsubsec:UnitScaling}

When cooling is neglected, Equations~(\ref{eq:CRMHD})-(\ref{eq:crmomentum}) can be non-dimensionalised and rescaled via three free parameters, which we will take to be the characteristic length scale $l_{0}$, temperature $T_{0}$, and density $\rho_{0}$.  The characteristic velocity $v_0$ is set to an isothermal sound speed using  $v_{0}\equiv c_{s,0}=\sqrt{k_{b}T_{0}/(0.6 m_{p})}$, where $k_{\rm B}$ is Botzmann's constant, $m_p$ is the proton mass.  This leads to dimensionless time $t_{0}=l_{0}/v_{0}$, energy density $E_{0}=\rho_0 v_0^2$, flux $F_0= \rho_0 v_0^3$, and acceleration $a_0=v_0^2/l_0$.  The choice of $T_0$, $\rho_0$, and $l_0$ can be rescaled to match varying galactic environments, as discussed in section~\ref{subsec:discussion_galaxies}.  The exception will be where cooling is discussed in section~\ref{subsec:result_cooling}, as cooling depends explicitly on the choice of temperature and density.

In the rest of paper, we report the dimensionless numbers unless otherwise noted. The fiducial vertical ($x$) and horizontal ($y$ or $z$) sizes are $L_{x}=400$ and $L_{y}=50$, with $8000\times1000$ total cells for the fiducial resolution. The aspect ratio of calculation domain varies between the simulations to adjust for specific problems as well as to minimize the computation cost. We discuss the effect of domain aspect ratio in Section~\ref{subsubsec:result_aspect_ratio}. In all simulations, we set the pressure floor $P_{\rm floor}=10^{-7}$ and density floor $\rho_{\rm floor}=10^{-8}$. We set the fiducial $\vm =1000$ and discuss the effect of $\vm$ in Section~\ref{subsubsec:result_resolution}.

\subsubsection{Initialization and boundary conditions}\label{subsubsec:InitializationBoundary}

We envision our simulation as an atmosphere situated above a star forming environment (e.g. a galactic disk).  The gravity $g$ is created by the gas and stars below the atmosphere and a CR flux is provided by the stellar winds and supernovae occurring in the (unsimulated) environment below.  Although idealised, this setup allows us to isolate and explore the impact of varying the CR flux in the launching region of the outflow, which is uncertain due to open questions about CR production, transport and destruction within the star forming environment.  The downside of this setup is that it does not admit steady state wind solutions of the form present in spherical symmetry or other geometry that allows the wind to expand into a larger cross-sectional area as it moves out \citep{ipavich1975galactic,mao2018galactic,quataert2021physics}.

We initialise the gas density to an isothermal profile in hydrostatic equilibrium with static background gravity. The gas density is uniform in the $y$ direction. In the $x$ direction, the distribution is the maximum of $\rho_0\exp(-x/h)$ and $\rho_{\rm bkgd}=7.988\times10^{-5}$.  We choose a scale height $h=2.018$.  The initial isothermal temperature $T_{\rm init}=1.667$.  Hydrostatic equilibrium for $\rho > \rho_{\rm bkgd}$ requires $g=-0.826$.  The magnetic field is initialised to be uniform with $B_{x}=B_{0}=2$ and all other components set to zero.

In the streaming limit, we set an isotropic diffusivity with $\kdiff=10^{-8}$. In the streaming-diffusion and non-streaming runs, we adopt isotropic diffusion with the values reported in Table~\ref{tab:HSE_summary_params}. The only exception is HSE\_1F\_hd\_aniso, where the diffusivity parallel to the field $\kappa_{\rm diff, \parallel}=100$ and the diffusivity perpendicular to the magnetic field $\kappa_{\rm diff, \bot}=10^{-8}$.

The boundaries are periodic in the horizontal ($y$ and $z$) directions. For the $x$ direction, the top MHD boundary is outflow, while the bottom MHD boundary is reflecting for hydro variables and outflow for magnetic field variables.  In order to prescribe a uniform CR flux, we fix the CR flux $F_{{\rm c},x}$ in the bottom ghost zones and copy the CR energy density, $y$ (and $z$ for three dimensional runs) component of the CR flux to the ghost zones from the last active zone. For the top CR boundary condition, we copy the CR flux and energy from the last active zone.  For the diffusion-only simulations, we implemented an alternate right MHD boundary that prevents any gas inflow, avoiding unphysical oscillations in the non-streaming simulations. We confirmed that the streaming runs were insensitive to this choice for the top MHD boundary condition.

Although we target a constant CR flux, we cannot precisely control the CR flux entering the domain when streaming dominates due to impacts from the CR ``bottleneck'' effect.  After an initial transient phase, the boundary settles to a nearly constant CR flux that is usually slightly below the target $F_{{\rm c}, x}$, with temporal variations of several percent or less of the average value. Higher target fluxes provide results closer to the target value. For the fiducial expected $F_{{\rm c},x}=15.0$, the time average between $5.0<t<25.0$ of CR flux across the boundary in the simulation is $\overline{F}_{\rm CR, in}=12.2$. We define the average difference
\begin{equation}
    \overline{\Delta F}_{\rm c,in}=\frac{\int_{t}\left|<\mathcal{F}_{E_{\rm c}}>-\bar{F}_{\rm c, in}\right|dt}{\int_{t}dt},\nonumber
\end{equation}
where $<X>=\int_{\rm in}XdA/\int_{\rm in}dA$ is volume average of $X$ at the bottom boundary, $\mathcal{F}_{E_{\rm c}}$ is the flux of CR energy returned by Riemann solver at the bottom boundary.  For a target flux $F_{{\rm c},x}=15.0$, the fluctuation $\overline{\Delta F}_{\rm c,in}/\overline{F}_{\rm c, in}=7.5\%$.  For target CR fluxes $F_{{\rm c},x}=60.0$ and $F_{{\rm c},x}=135.0$, the average CR flux at the boundary is $\overline{F}_{\rm c,in}=59.074$, $\overline{F}_{\rm c,in}=134.451$, and fluctuation are only $1.1\%$ and $0.5\%$ accordingly.

\section{Results}\label{sec:results}

\begin{table*}
\caption{Summary of Simulation Parameters}
\label{tab:HSE_summary_params}
\begin{threeparttable}
\centering
\begin{tabular}{lcccccc}
\hline
Name & $\kappa$ & $F_{\rm c}$ & $B_{0}$  & $L_{x}\times L_{y}$ \tnote{b} & $N_{x}\times N_{z}$ \tnote{b} & $\vm$\\
\hline
HSE\_1F\_str & $10^{-8}$ & 15.0 & 2.0 & $400\times50$ & $8000\times1000$ & 1000 \\
HSE\_hF\_str & $10^{-8}$ & 7.5 & 2.0 & $400\times50$ & $8000\times1000$ & 1000 \\
HSE\_4F\_str & $10^{-8}$ & 60.0 & 2.0 & $800\times50$ & $16000\times1000$ & 1000 \\
HSE\_9F\_str & $10^{-8}$ & 135.0 & 2.0 & $800\times50$ & $16000\times1000$ & 1000 \\
HSE\_20F\_str & $10^{-8}$ & 300.0 & 2.0 & $800\times50$ & $16000\times1000$ & 1000 \\
HSE\_1F\_ld & $1.0$ & 15.0 & 2.0 & $200\times5$ & $4000\times100$ & 1000 \\
HSE\_1F\_cd & $10.0$ & 15.0 & 2.0 & $200\times25$ & $4000\times500$ & 1000 \\
HSE\_1F\_hd & $100.0$ & 15.0 & 2.0 & $200\times5$ & $4000\times100$ & 1000 \\
HSE\_1F\_hd\_aniso & $100.0$ & 15.0 & 2.0 & $200\times5$ & $4000\times100$ & 1000 \\
HSE\_1F\_hd\_ns\tnote{a} & $100.0$ & 15.0 & 2.0 & $200\times25$ & $4000\times500$ & 1000 \\
HSE\_1F\_ld\_ns\tnote{a} & $1.0$ & 15.0 & 2.0 & $200\times25$ & $4000\times500$ & 1000 \\
HSE\_1F\_str\_sc\tnote{c} & $10^{-8}$ & 15.0 & 2.0 & $400\times50$ & $8000\times1000$ & 1000 \\
HSE\_1F\_str\_wc\tnote{c}& $10^{-8}$ & 15.0 & 2.0 & $400\times50$ & $8000\times1000$ & 1000 \\
HSE\_9F\_str\_sc\tnote{c} & $10^{-8}$ & 135.0 & 2.0 & $400\times50$ & $8000\times1000$ & 1000 \\
HSE\_1F\_str\_b1 & $10^{-8}$ & 15.0 & 1.0 & $200\times5$ & $4000\times100$ & 1000 \\
HSE\_1F\_str\_b4 & $10^{-8}$ & 15.0 & 4.0 & $200\times5$ & $4000\times100$ & 1000 \\
\hline
HSE\_1F\_str\_VM & $100.0$ & 15.0 & 2.0 & $400\times50$ & $8000\times1000$ & 2000 \\
HSE\_1F\_str\_LR & $10^{-8}$ & 15.0 & 2.0 & $400\times50$ & $4000\times500$ & 1000 \\
HSE\_1F\_str\_HR & $10^{-8}$ & 15.0 & 2.0 & $400\times50$ & $16000\times2000$ & 1000 \\
HSE\_1F\_str\_3D & $10^{-8}$ & 15.0 & 2.0 & $400\times25\times25$ & $4000\times250\times250$ & 1000 \\
\hline

\end{tabular}
\begin{tablenotes}
    \item[a]Streaming is turned off, equivalently setting $\vs=0$ in Equation~\ref{eq:gasenergy}, Equation~\ref{eq:crenergy}, $\vel_{\rm A}=0$ in Equation~\ref{eq:sigma}.
    \item[b]The reported dimension corresponds to the largest simulation domain for each set of parameters. Notice that some of the simulation has lower aspect ratio $L_{y}/L_{x}$, we discuss the effect of aspect ratio in Section~\ref{subsubsec:result_aspect_ratio}
    \item[c]The three adiabatic simulations with different cooling strength are scaled to different physical units.
\end{tablenotes}
\end{threeparttable}
\end{table*}

\subsection{The Effect of CR Flux on Streaming: from sub-Eddington to super-Eddington }\label{subsec:result_stream}

The five streaming dominated simulations are HSE\_hF\_str HSE\_1F\_str, HSE\_4F\_str, HSE\_9F\_str and HSE\_20F\_str, the diffusivity is set to be $\kappa_{\rm diff}=10^{-8}$, so that $\sigma_{\rm c}^{-1}\approx\frac{\textbf{B}}{|\textbf{B}\cdot(\nabla\cdot\textsf{P}_{\rm c})|}\vel_{\rm A}\cdot(E_{\rm c}\textsf{I}+\textsf{P}_{\rm c})$. We estimate the Eddington flux according to Equation~\ref{eq:feddstreaming}. Assuming $v_{\rm s}$ is calculated using the initial density $\rho_{\rm init}=\rho_{0}$ at the base of the atmosphere and the initial uniform magnetic field $B_{0}$, the estimated Eddington flux $F_{\rm edd, str}=20.0$.

We select HSE\_1F\_str with CR flux $F_{\rm c}=15.0$ as our fiducial run. This is somewhat sub-Eddington at the base of the domain, but becomes super-Eddington as the density drops. 

\begin{figure}
    \centering
    \includegraphics[width=\columnwidth]{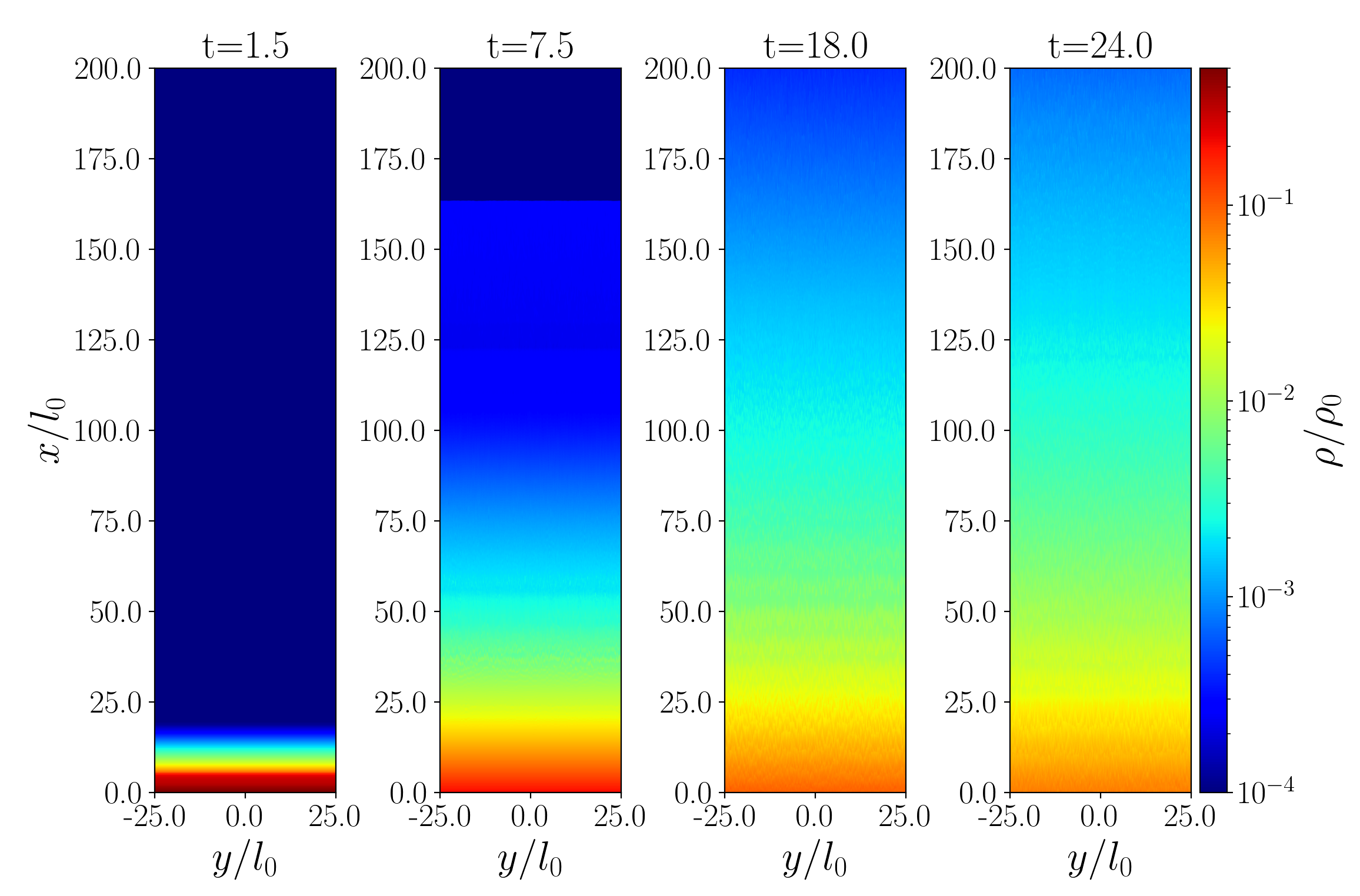}
    \caption{The density snapshots of HSE\_1F\_str. From left to right, t=1.5, 7.5, 18.0 and 24.0. The picture shows only half of the simulation domain in the $x$ direction.}
    \label{fig:1Fstr_density}
\end{figure}

\begin{figure}
    \centering
    \includegraphics[width=\linewidth]{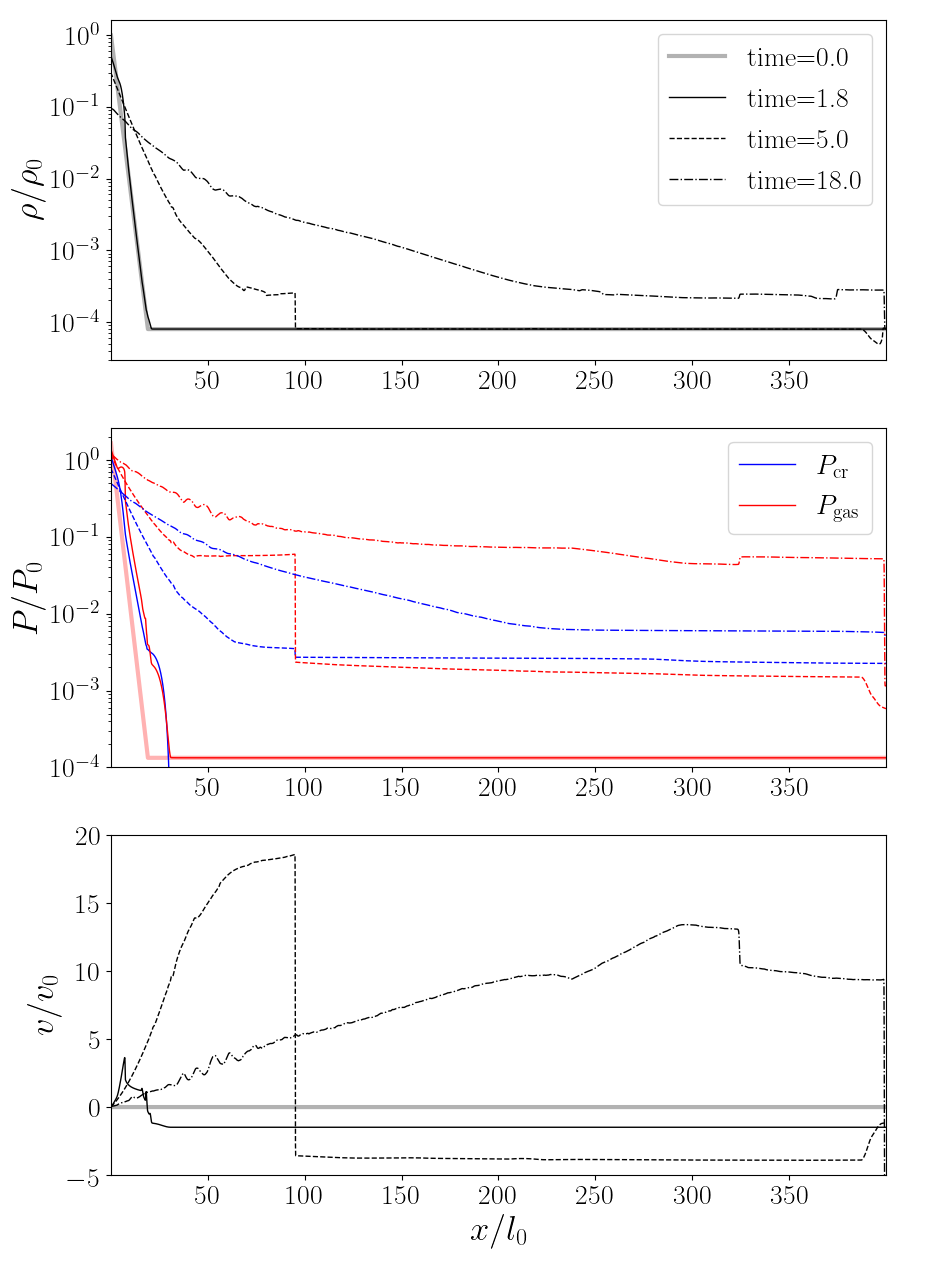}
    \caption{The $y$ direction averaged density (the first row), CR pressure (the blue lines in the second row), gas pressure (the red lines in the second row) and gas velocity (the third row) of HSE\_1F\_str. The line style corresponds to different time, the thick solid lines in each panel are the initial profile. Other sampled times are t=1.8 (the thin solid lines), t=5.0 (the dashed lines) and t=18.0 (the dashed dotted lines).}
    \label{fig:1Fstrlineplot}
\end{figure}

Figure~\ref{fig:1Fstr_density} shows density snapshots from HSE\_1F\_str for the bottom half of the domain. At early time $t=1.5$, while the CR is streaming in the dense atmosphere, the gas density profile remains close to the initial distribution.  This is shorter than the timescale for CRs to stream out of the gas, which we approximate as the Alfv\'{e}n-wave crossing time:
\begin{eqnarray}\label{eq:tstr}
    t_{\rm str}=\int_{0}^{2l_{0}}\frac{dx}{v_{\rm s}(x)}\nonumber
    &=\frac{2h_{0}}{B_{0}/\sqrt{\rho_{0}}}(1-e^{-\frac{l_{0}}{h_{0}}}),
\end{eqnarray}
which gives $t_{\rm str}\sim 2.0$ for fiducial values.  At later times, we see that a substantial fraction of the gas is lofted up as the atmosphere expands outward under a combination of CR and gas pressure.  Since mass is conserved, the density drops near the base.  Snapshots at $t=18.0$ and $t=24.0$ are similar, suggesting the gas is reaching a quasi equilibrium density profile, even though a modest outflow is present at these later times. There is also evidence of abrupt jumps in the density that show up more clearly in Figure~\ref{fig:1Fstrlineplot}.  We attribute these inhomogeneities to the development of an acoustic instability, as discussed in section~\ref{subsec:discussion_ins}. Figure~\ref{fig:1Fstr_density} shows that this instability does not lead to large amplitude variations on the horizontal dimensions, at least for the magnetic field strength and geometry used here.

Figure~\ref{fig:1Fstrlineplot} show horizontally averaged quantities and illustrates the ``bottleneck'' effect. When the CR force acts on gas and streaming dominates the CR flux, the streaming velocity $v_{\rm s}\propto1/\sqrt{\rho}$, increasing as the gas density drops. When CRs stream into the atmosphere, CRs tend to pile up in the dense region where they stream slower, resulting in a CR pressure gradient from dense gas to diffuse gas, producing the ``bottleneck''. Given the monotonic density profile, $\pcr$ reaches its maximum at the base of the atmosphere and decreases outward. 

Around $t=1.8$, the gas density is similar to the initial profile and nearly unperturbed. The gas pressure closely follows the CR pressure, and the combined pressure gradients exceed the gravitational force, launching the atmosphere upward. By $t=5.0$, the CRs begin to stream out of the dense gas and drive the expansion, enhancing both the CR and gas pressure in the initially low density background region ($x\gtrsim 20$).  The large enhancement in gas pressure is driven by CR heating. Gas velocities are quite high for $x < 100 l_0$, due to the combined acceleration by CR and gas pressure gradients.

As more gas is pushed towards the upper boundary, the density gradient becomes shallower and the Alfv\'{e}n velocity gradient consequently decreases (the $B$ field remains more constant). Therefore, the CR streaming ``bottleneck'' is less effective, the CR pressure gradient decreases, providing less heating and acceleration. The velocity decreases as more mass is swept up in the outflow and continues to do work against the constant gravitational acceleration.  The combined gas and CR pressure gradients still exceed gravity, with velocity increasing upward due to the resulting acceleration. At $t=18.0$, the diffuse gas starts to flow out of the domain, carrying away a modest faction of the initial mass. After this point and before we end the simulation at $t=25.0$, all the quantities shown in this plot are changing slowly, with pressure and density gradients gradually becoming more shallow with time.

Even after horizontal averaging, the effects of the instabilities noted above can be discerned as the growth of a series of shocks in the gas density, pressure, and velocity profiles for $45\lesssim x\lesssim 75$ at $t=18.0$. We identify them as the result of CR acoustic instability and discuss them in more detail in Section~\ref{subsec:discussion_ins}. The instability, however, saturates at a level where it does not significantly change the volume averaged CR force. Despite the instability, the magnetic field is relatively unperturbed, making CRs effectively stream along the uniform magnetic field. Hence, there is no analogue of the Rayleigh-Taylor instability that regulates the acceleration of gas in radiatively driven atmospheres \citep{2012ApJ...760..155K}.

The overall dynamical picture is that CRs accelerate gas against gravity even with this moderate CR flux, but a significant fraction of the energy goes into heating. According to Equation~\ref{eq:gasenergy}, CRs heat the gas at the rate of $\va\cdot\nabla\pcr$. Since $v_{\rm A}\propto1/\sqrt{\rho}$, the lower density regions tend to experience greater heating.  Gas is in the low density background can be heated by over two orders of magnitude with the lowest density gas experiencing the largest heating.  These results suggest radiative cooling could play a significant role and we consider its impact below.

\begin{figure}
    \centering
    \includegraphics[width=\linewidth]{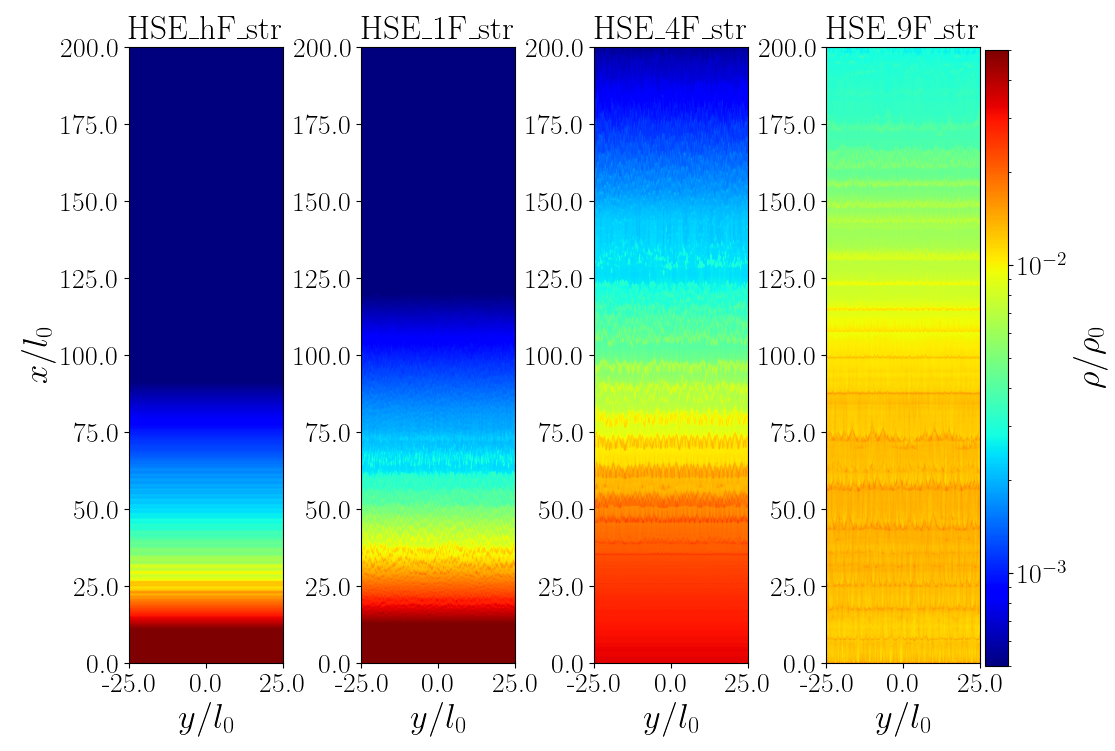}
    \caption{The density snapshots at $t=10.0$ for four streaming dominated simulations with different CR flux. From left to right, the injected CR flux $F_{\rm c}=7.5$(HSE\_hF\_str), 15.0(HSE\_1F\_str), 60.0(HSE\_4F\_str), 135.0(HSE\_9F\_str). We only show part of the domain in the $x$ direction (see $L_{x}$ in Table~\ref{tab:HSE_summary_params}).}
    \label{fig:density_flux}
\end{figure}

\begin{figure}
    \centering
    \includegraphics[width=\linewidth]{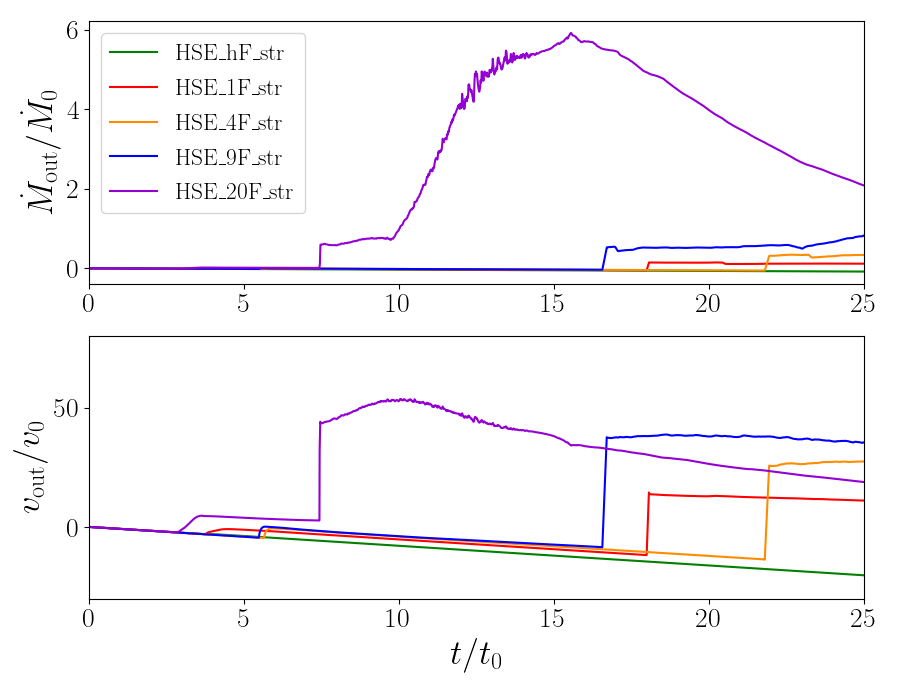}
    \caption{The gas outflow from simulations in the streaming limit with different injecting CR flux: HSE\_hF\_str (green), HSE\_1F\_str (red), HSE\_4F\_str (red), HSE\_9F\_str (blue) and HSE\_20F\_str (purple), measured at $x=400l_{0}$. In this plot, we report the results from simulations with the same box size $L_{x}=400$ instead of the reported $L_{x}$ in Table~\ref{tab:HSE_summary_params}. Upper panel: the mass outflow rate per unit area. Lower panel: the average outflow velocity. Both defined as Equation~\ref{eq:massoutflow}}
    \label{fig:outflow_flux}
\end{figure}

Figures~\ref{fig:density_flux} and \ref{fig:outflow_flux} show the impact of varying the target CR flux injected at the base of the atmosphere.  We consider simulations with CR fluxes 0.5 (HSE\_hF\_str), 4 (HSE\_4F\_str), 9 (HSE\_9F\_str) and 20 (HSE\_20F\_str) times the flux in the fiducial run. Although the overall dynamics are similar, HSE\_4F\_str and HSE\_9F\_str produce more significant gas outflow in shorter time. In contrast, in the sub-Eddington case HSE\_hf\_str with half of the fiducial CR flux, the atmosphere expands outward over a longer timescale, but by the time we finish other simulations at $t=25$, there is no significant gas outflow. Figure~\ref{fig:density_flux} compares the gas density snapshot at $t=10.0$ of these simulations. As we increase the CR flux from sub-Eddington to super-Eddington, the atmosphere is more dispersed and the outflow carries larger mass and momentum. We see that as the CR flux increases, the shocks driven by acoustic instability become more prominent, but the amplitude never becomes large enough to have a significant feedback on the gas acceleration.

Figure~\ref{fig:outflow_flux} shows the volume average of the mass outflow rate per area $\dot{M}_{\rm out}$ (the upper panel) and averaged outflow velocity $v_{\rm out}$ (the lower panel) at $x=400 l_0$ as a function of time. $\dot{M}_{\rm out}$ and $v_{\rm out}$ are calculated as:
\begin{equation}\label{eq:massoutflow}
    \dot{M}_{\rm out}=\frac{\int_{t}\int_{\rm out}\mathcal{F}_{\rho}dydt}{\int_{t}\int_{\rm out}dydt},\quad
    v_{\rm out}=\frac{\int_{t}\int_{\rm out}\rho vvdxdydt}{\int_{t}\int_{\rm out}\mathcal{F}_{\rho}dydt},
\end{equation}
where $\mathcal{F}_{\rho}$ is the density flux returned by Riemann solver on the outer boundary, $\rho vv|_{\rm out}$ is the $\rho vv$ calculated at the outer boundary, $\int_{\rm out}$ denotes integration on the outer boundary.  At early times, there is modest inflow and velocities are negative due to the fact that the isothermal and constant density background has no pressure gradient to support the gas against gravity.  For the sub-Eddington run, the combination of CR and gas pressure is never sufficient to overcome gravity and this inflow persists to the end of the simulation.  For the marginal and super-Eddington cases, the CR and gas pressure forces eventually exceed gravity and drive outflows.  The general trend is that both the velocities and mass outflow rates increase as the injected CR flux increases.

\begin{figure}
    \centering
    \includegraphics[width=0.5\textwidth]{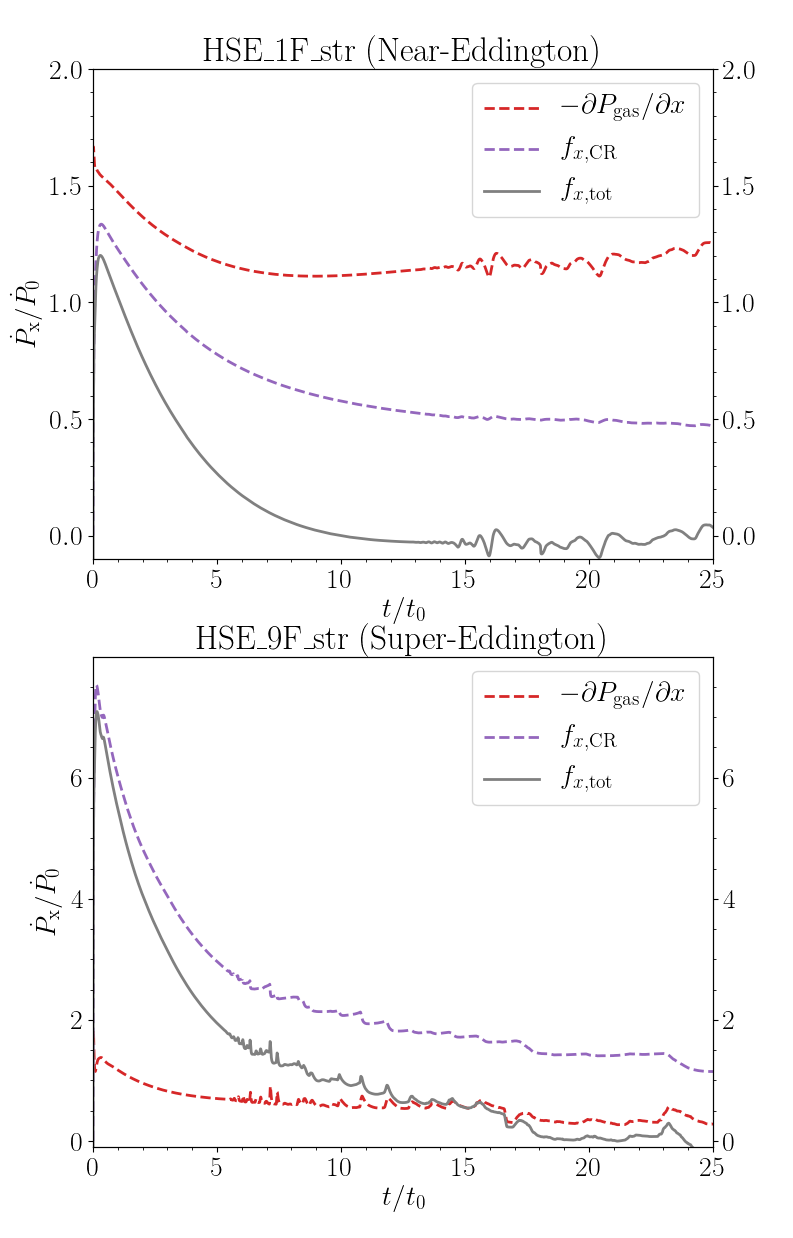}
    \caption{Force terms in the $x$ direction as a function of time for the nearly-Eddington (the top panel) HSE\_1F\_str and the Super-Eddington case HSE\_9F\_str (the bottom panel). The force terms are summed over across the domain. In both panels, the red dashed line is gas pressure gradient force, the purple dashed line is CR pressure gradient force, and the gray solid line is total $x$ direction force. The rate is sampled every $\Delta t=0.001$.}
    \label{fig:gasrate_flux}
\end{figure}

The outflow rates in simulations reflect the different force balance in sub-Eddington systems and super-Eddington systems. Figure~\ref{fig:gasrate_flux} shows the $x$ direction forces as a function of time for the fiducial near-Eddington case (HSE\_1F\_str) and a super-Eddington case (HSE\_9F\_str). Here we only explicitly plot the gas pressure gradient force (the red dashed line) and CR pressure gradient force (the purple dashed line), the sum with other forces including gravity, magnetic field force and advected momentum is the grey solid line. A direct difference between near-Eddington and super-Eddington is the inversion of CR and gas pressure gradient force. We also examine other simulations with different CR flux, the super-Eddington runs always have a higher CR force than gas pressure gradient force, and vice versa for the sub-Eddington and near-Eddington cases. 

Despite the different interplay between gas pressure and CR forces, the CR force (dashed purple lines) follows similar time evolution: it peaks at the beginning, then relaxes to a lower value in a few $t_{\rm str}$, and slowly decreases. This is because the ``bottleneck'' effect relies on gas density gradient. However, as gas is accelerated and dispersed, the density gradient becomes shallower and even vanishes, so the CR pressure gradient forces drops commensurately at later time. The gas pressure gradient is modified by CR-gas interactions, and contributes to support against gravity. Although the total force tends to be relatively constant and close to zero at later times, there is no steady state wind developed in our simulations since the initial gas is redistributed and slowly depleted. Nevertheless, our results confirm that the CR Eddington flux is a good criterion for distinguishing the presence of significant levels of CR driven outflow.

\subsection{The Effect of CR Transport Mechanism: Streaming versus Diffusion}

In this section, we consider the relative importance of CR diffusion and streaming. Although all the streaming runs reported above have a small nominal diffusive component ($\kdiff=10^{-8}$), the CR flux is completely dominated by streaming.  We first consider three hybrid cases where both streaming and diffusion are present: low diffusivity (HSE\_1F\_ld $\kdiff=1.0$), comparable diffusivity (HSE\_1F\_cd, $\kdiff=10.0$), and high diffusivity (HSE\_1F\_hd, $\kdiff=100.0$) but parameters are otherwise set to match the fiducial run (HSE\_1F\_str). We also consider two diffusion (non-streaming) simulations with high diffusivity (HSE\_1F\_hd\_ns, $\kdiff=100.0$) and low diffusivity (HSE\_1F\_ld\_ns, $\kdiff=1.0$). These non-streaming simulations have $\vs=0$ and $\va=0$ in Equation~\ref{eq:CRMHD}-\ref{eq:crmomentum} and Equation~\ref{eq:sigma}, so CRs cannot heat the gas via wave damping.

\begin{figure}
    \centering
    \includegraphics[width=\linewidth]{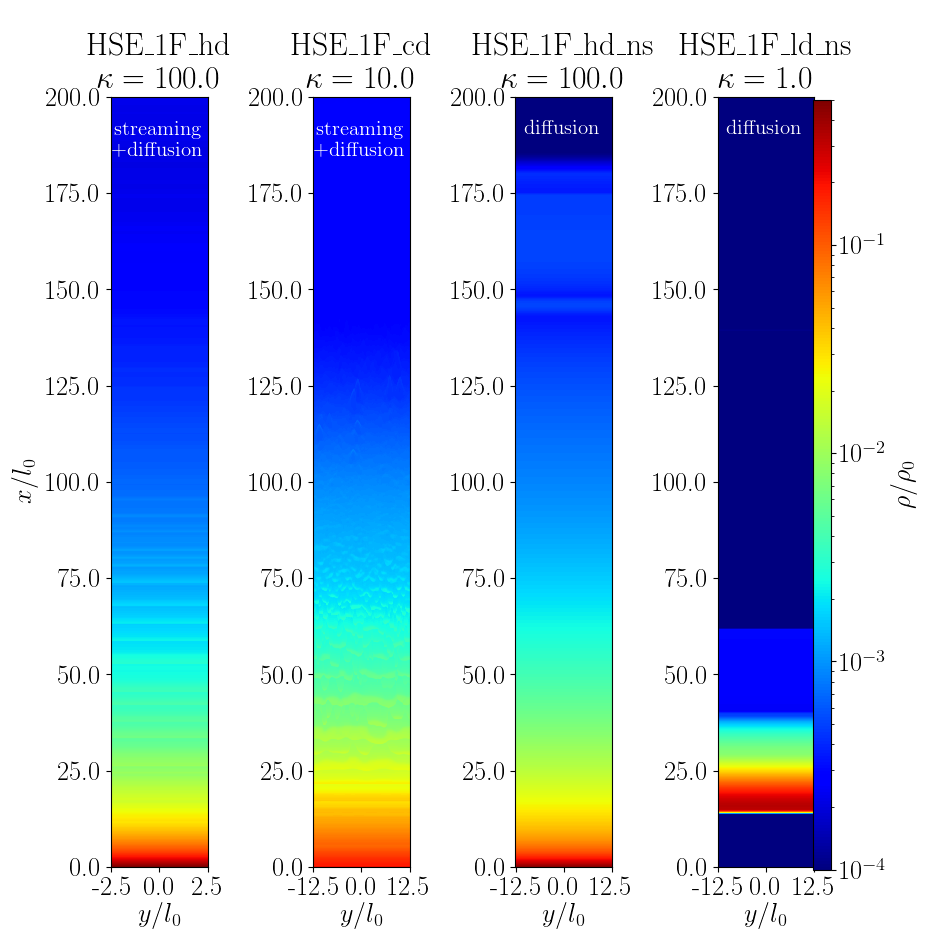}
    \caption{The density snapshots at $t=10.0$ for different CR transport models. From left to right, the first two columns are streaming and diffusion, the diffusivity is $\kdiff=10.0$(HSE\_1F\_cd) and $\kdiff=100.0$ (HSE\_1F\_hd). The thrid and fourth columns are pure diffusion without streaming. The diffusivity are accordingly $\kdiff=100.0$ (HSE\_1F\_hd\_ns) and $\kdiff=1.0$ (HSE\_1F\_ld\_ns). The aspect ratio of plots are adjusted for the purpose of visual comparison with each other, the Y labels indicate the actual size of calculation domain.}
    \label{fig:density_diff}
\end{figure}

We first look at the diffusion-only simulations. The CR flux $\Fc\sim-\kdiff\nabla\pcr$, so $\nabla\pcr\sim-\Fc/\kdiff$. If we ignore the time-dependent term in Equation~(\ref{eq:crmomentum}), the CR contribution to gas momentum at RHS of Equation~(\ref{eq:gasmomentum}) is $\nabla\pcr$. So for a given CR flux, larger $\kdiff$ gives the smaller acceleration since CRs diffuse through the gas more quickly without much interaction. The gas density for HSE\_1F\_hd\_ns and HSE\_1F\_ld\_ns at $t=10.0$ is shown in the third and fourth column in Figure~(\ref{fig:density_diff}) respectively. In HSE\_1F\_ld\_ns CRs cannot diffuse through the gas rapidly, so a strong CR pressure gradient develops at the base of the atmosphere, with the CR force exceeding gravity and lofting the entire atmosphere.  Presumably, there would be feedback on the injected CR flux in a more realistic set-up but the key result here is that CRs are capable of collectively launching and accelerating the gas against gravity, with no instabilities (e.g. Parker or Rayleigh-Taylor like) present to disrupt the shell.

In contrast, HSE\_1F\_hd\_ns experiences a more gentle acceleration. CRs diffuse through the gas expanding the atmosphere in a manner more similar to the streaming runs but with very little outflow. This difference between $\kdiff=100.0$ and $\kdiff=1.0$ dynamics also can be interpreted in terms of the CR Eddington flux. For the injected CR flux $F_{\rm c}\approx15.0$, the low diffusivity ($\kdiff=1.0$) run is super-Eddington at the base ($F_{\rm edd}\approx0.83$), but the high diffusivity ($\kdiff=100.0$) run is substantially sub-Eddington ($F_{\rm edd}\approx82.62$) so the atmosphere relaxes to steady state with a larger scale height rather than driving outflow.

We turn on both streaming and diffusion in HSE\_1F\_ld, HSE\_1F\_cd and HSE\_1F\_hd. The gas momentum source term (RHS of Equation~\ref{eq:gasmomentum}) is proportional to the interaction coefficient $\sigma_{\rm c}$. The first term in the interaction coefficient is the classical diffusivity, the second term is $\sigma_{\rm str}^{-1}$ is an effective ``diffusivity'' for streaming. We estimate it based on the initial value at the base of atmosphere: $\sigma_{\rm str,init}^{-1}\approx4v_{\rm A}/(\partial\ln P_{\rm c}/\partial x)\approx-6v_{\rm A}H_{\rho}\approx24.2$ (see similar derivation in Equation~\ref{eq:feddstreaming}). In HSE\_1F\_ld we set $\kdiff=\sigma_{\rm c}'^{-1}=1.0$, so that the interaction coefficient is dominated by the effective streaming coefficient, while in HSE\_1F\_hd $\kdiff=100.0$, so the diffusivity dominates. The isotropic diffusivity in HSE\_1F\_cd is $\kdiff=10.0$, which is about the same order of magnitude as $\sigma_{\rm str,init}^{-1}$.

The first two columns in Figure~\ref{fig:density_diff} compare the density snapshots for the streaming-diffusion simulations:  HSE\_1F\_cd and HSE\_1F\_hd at $t=10.0$.  The gas distribution in HSE\_1F\_ld (not shown) is almost identical to the streaming simulation HSE\_1F\_str with the same CR flux shown in Figure~\ref{fig:1Fstr_density}. In HSE\_1F\_ld, the small diffusivity does not lead to a large CR pressure gradient as the pure diffusion case because $\sigma_{\rm c}^{-1}$ is now dominated by the streaming component.  By preventing the development of large gradients, CR streaming limits the CR force when CR diffusivity is low. At the other end where the diffusivity is large, the gas density profile in HSE\_1F\_hd is similar to the non-streaming case HSE\_1F\_hd\_ns with the same diffusivity, except for the presence of shocks formed from the acoustic instability. In HSE\_1F\_cd, where streaming and diffusion are comparable, the result falls between these limits, but most similar to the  HSE\_1F\_ld run.

\begin{figure}
    \centering
    \includegraphics[width=1.0\linewidth]{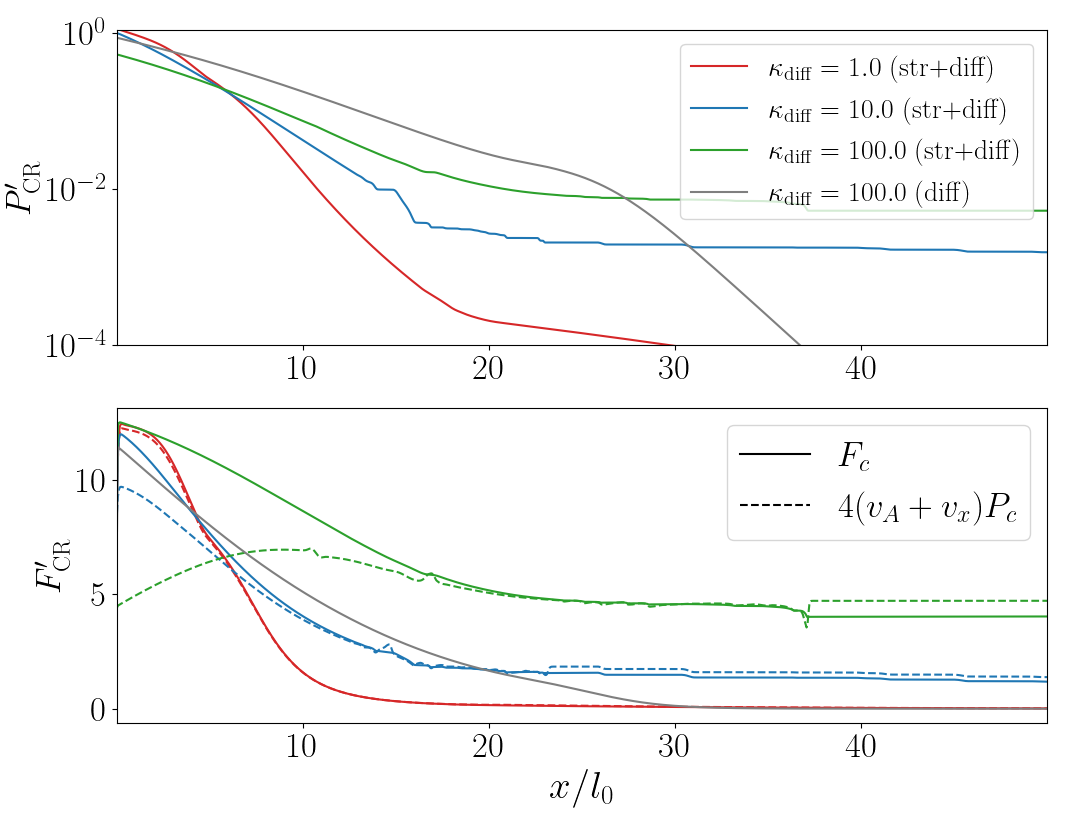}
    \caption{$y$ direction averaged CR pressure (the first row), total CR flux (solid lines in the second row), the streaming flux component (dashed lines in the second row) for $0<x<50$ at $t=1.5$ for the diffusion runs: HSE\_1F\_hd (grey solid line) and the three streaming-diffusion simulations: HSE\_1F\_str\_ld (red lines), HSE\_1F\_str\_cd (blue lines), HSE\_1F\_str\_hd (green lines). Note that there is no grey dashed curve because streaming is not relevant here.}
    \label{fig:cr_relation}
\end{figure}

When $\kdiff$ is small, the interaction coefficient is dominated by the streaming component, and the dynamics is similar to streaming limit. Increasing $\kdiff$ will ensure that $\sigma_{c}^{-1}\sim\kdiff$ but also lower the interaction coefficient, which usually leads to a reduction of the CR pressure gradient, reducing the CR force on the gas. The upper panel of Figure~\ref{fig:cr_relation} illustrates the decrease in the CR pressure gradient as $\kdiff$ increases in the streaming-diffusion runs at early time ($t=1.5$), approximately when the CR force peaks. 

The bottom panel shows the total CR flux (the solid lines) and the streaming flux (the dashed line) at the same time. Since gas has not yet received a significant acceleration, the advection flux ($4v_{\rm x}P_{\rm c}$) is small. Thus, the difference between total CR flux and streaming flux is dominated by the diffusive flux. When diffusivity is low, the CR flux is dominated by the streaming flux ($4v_{\rm A}P_{\rm CR}$), but the diffusive flux is significant wherever the pressure gradient is present and diffusivity is large. To illustrate the difference, we also show the CR flux for the diffusion-only run with $\kappa_{\rm{diff}}=100.0$ as the grey solid line, which does not have the streaming component. Although changing $\kdiff$ modifies the balance of streaming and diffusive flux, the total flux is fixed by our boundary condition, so $\sigma_{\rm c}$ roughly controls the amount of momentum that CR transfers to gas.  The upshot is that for a given CR flux, the overall momentum transfer from CRs to gas is generally reduced when streaming operates.

\subsection{The Impact of Cooling}\label{subsec:result_cooling}

In this section, we study the effect of radiative cooling by comparing three streaming-dominated simulations with different cooling times.  A more detailed treatment of cooling and the effects of thermal instability on CR driven outflows will be presented in a companion paper (Huang, Jiang \& Davis, in preparation). We adopt an optically thin radiative model with cooling rate  $n_{H}^{2}\Lambda(T)$ and  a constant supplemental heating rate $n_{H}\Gamma$. The cooling function $\Lambda(T)$ is taken from \citet{2019MNRAS.489..205W}, which is an approximation to fit the CLOUDY data from \citet{wiersma2009effect} assuming solar metallicity:
\begin{equation}
    \Lambda(T)=1.1\times10^{-21}\times 10^{\Theta(\log(T/10^{5}\rm K))}\rm erg~cm^{3}~s^{-1},
\end{equation}
where 
\begin{equation}
    \Theta(x)=0.4x-3+\frac{5.2}{e^{x+0.08}+e^{-1.5(x+0.08)}}
\end{equation}
We cut off the radiative cooling for gas with temperature below $10^{4}$K.

Because cooling is proportional to $n_{H}^{2}$, the cooling time significantly increases as the gas density decreases with height. The initial cooling timescale at two scale heights is larger than the cooling timescale at the base by more than two orders of magnitude. We report our simulations in terms of $t_{\rm cool}/t_{0}$, the ratio of the cooling time to sound crossing time at the base of the simulation, but this will overestimate the importance of cooling higher in the simulation where densities are lower. If $t_{\rm cool} < t_0$ we expect cooling to impact the gas pressure gradient, but if $t_{\rm cool} >  t_{0}$, cooling may not be important for the gas dynamics. 

In contrast to our non-radiative simulations, introduction of cooling fixes the density and temperature of the simulation, removing the scaling freedom that was previously present. 
HSE\_1F\_str\_wc has $t_{\rm cool}/t_{0}=0.7$ (initially) at the base. The density scaling unit is $\rho_{0}=2.35\times10^{-27}\rm g~cm^{-3}$, the length unit is $l_{0}=1.54\times10^{20}\rm cm$, and the temperature unit $T_{0}=9.9\times10^{4}\rm K$. For HSE\_1F\_str\_sc and HSE\_9F\_str\_sc the density scaling unit is $\rho_{0}=2.35\times10^{-25}\rm g~cm^{-3}$, the length unit is $l_{0}=1.54\times10^{20}\rm cm$, and the temperature unit $T_{0}=9.9\times10^{3}\rm K$, yielding $t_{\rm cool}/t_{0}=0.06$ at the base. However, we continue to report results in dimensionless numbers to facilitate comparison to non-radiative simulations.

\begin{figure}
    \centering
    \includegraphics[width=0.8\linewidth]{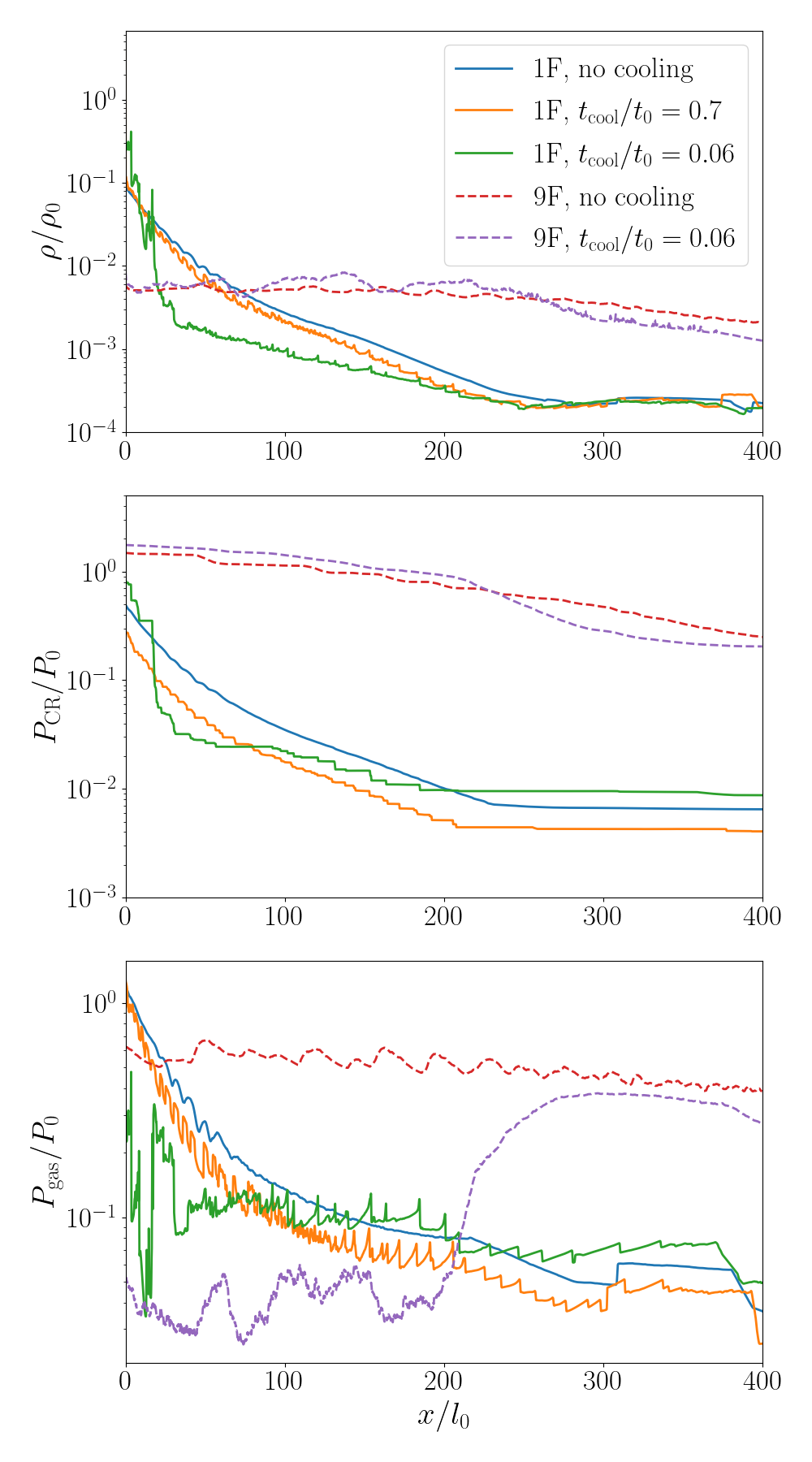}
    \caption{The $y$ direction averaged gas density (the first row), gas pressure (the second row), CR pressure (the third row) for the marginal-Eddington flux: HSE\_1F\_str (blue solid lines, no cooling), HSE\_1F\_str\_wc (orange solid lines, weak cooling) and HSE\_1F\_str\_sc (green solid lines, strong cooling), and the super-Eddington flux HSE\_9F\_str (red dashed lines, no cooling) and HSE\_9F\_str\_sc (purple dashed lines, strong cooling) The snapshots are sampled at $t=20$. }
    \label{fig:rhopress_cool}
\end{figure}

\begin{figure}
    \centering
    \includegraphics[width=\linewidth]{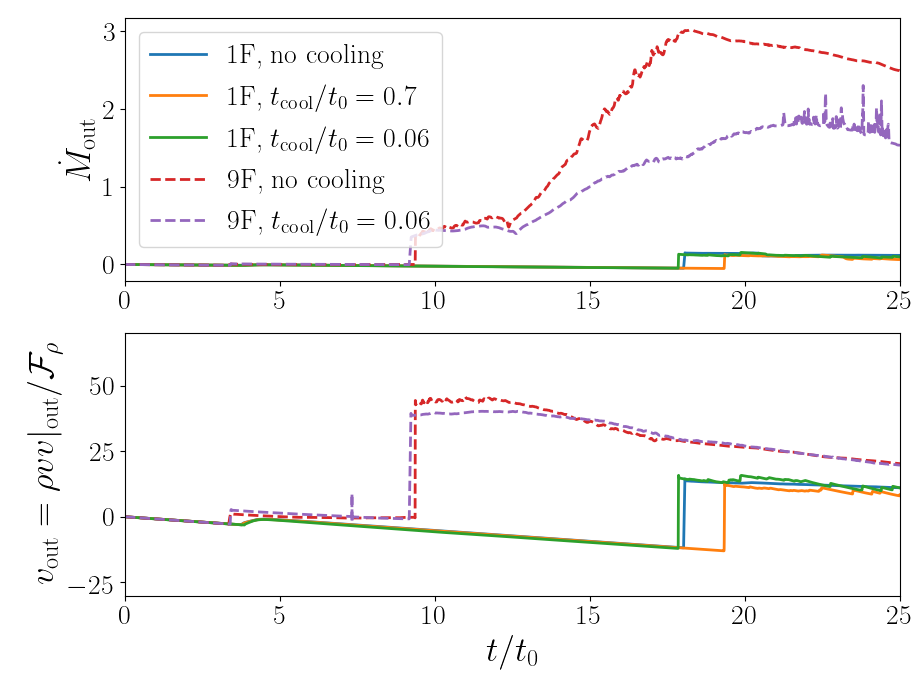}
    \caption{The mass outflow rate per unit area (the upper panel) and average velocity (the lower panel) for the marginal-Eddington flux: HSE\_1F\_str (blue solid lines, no cooling), HSE\_1F\_str\_wc (orange solid lines, weak cooling) and HSE\_1F\_str\_sc (green solid lines, strong cooling), and super-Eddington flux HSE\_9F\_str (red dashed lines, no cooling) and HSE\_9F\_str\_sc (purple dashed lines, strong cooling). Comparing the simulations without cooling, with short cooling timescale and long cooling timescale, the three marginal-Eddington runs shows similar outflow properties, the two super-Eddington runs have different mass outflow rate, but similar velocity. }
    \label{fig:outflow_cool}
\end{figure}

Figure~\ref{fig:rhopress_cool} shows snapshots of horizontally ($y$) averaged gas density (top panel), CR pressure (middle panel) and gas pressure (bottom panel) at $t'=20$. The blue solid line shows the profile of the non-radiative run HSE\_1F\_str, and the other two solid lines are the radiative simulations. The red dashed lines shows the non-radiative super-Eddington flux simulation HSE\_9F\_str, and the purple dashed line is the corresponding radiative simulation: HSE\_9F\_str\_sc.

For simulations with near-Eddington CR flux the orange lines show the profiles from the radiative run HSE\_1F\_str\_wc. The gas pressure at the base is lower due to the cooling, but because the cooling is relatively moderate, it has limited effect on the overall dynamics. The gas density and CR pressure are similar to non-radiative simulations. 

The green lines are the profiles from HSE\_1F\_str\_sc, where the short cooling time significantly alters the gas profile at the base compared to HSE\_1F\_str. Since the internal energy is radiated away, the gas lacks pressure support and falls toward the base, forming a much sharper density gradient and is less dispersed compared to the non-radiative and moderate cooling runs. The sharper density gradient increases the bottleneck effect for CR streaming, leading to a stronger CR force on the gas.

The horizontally averaged gas density profile is more strongly clumped in both radiative runs. The shocks created by acoustic instability are enhanced by cooling. Since the shocks are relatively isothermal, the enhanced density in the shock increases the cooling rate, lowering gas pressure relative to the shocks in non-radiative runs. The relatively lower pressure support allows the gas to condense more effectively.

Despite the differences in the gas density profiles, the outflow properties from the three near-Eddington runs shows in Figure~\ref{fig:outflow_cool} are similar. The outflow we observed in the simulations with near-Eddington CR flux are composed primarily of low density gas, where the cooling time is significantly longer than dynamical timescales. An interesting contrast is provided the super-Eddington simulation HSE\_9F\_str\_sc, which also has a short cooling time and corresponds to the purple dashed line in Figure~\ref{fig:rhopress_cool} and Figure~\ref{fig:outflow_cool}. 

In the lower panel of Figure~\ref{fig:rhopress_cool}, the gas pressure is significantly reduced by cooling and the gas pressure gradient is inverted compared to the initial profile. The gas density profile is similar to its non-radiative counterpart simulation HSE\_9F\_str (red dashed line), however, because the CR force dominates over the gas pressure force. The momentum transfer between CRs and gas that primarily determines the gas density and velocity distribution in super-Eddington systems is only indirectly impacted by radiative cooling. Figure~\ref{fig:outflow_cool} shows that the mass outflow rates are reduced by cooling but the effect is less than 50\% and velocities are relatively unimpacted. The upshot is that cooling is likely to be important in setting density profiles in the wind launching regions, but is probably less important in the super-Eddington cases, where CR forces are more important than cooling.  Cooling may still be very important in the subsequent outflow dynamics \citep[Huang, Jiang \& Davis, in prep.]{wiener2017cosmic,bustard2020cosmic,hopkins2020but} or in the overall structure of the CGM \citep{ji2020properties}.

\subsection{The Effect of Simulation Parameters}

In this section, we  briefly discuss a number of factors in our setup and their impact on the numerical results, including the simulation dimensions and aspect ratio in Section~\ref{subsubsec:result_aspect_ratio}, along with resolution, dimensionality, and choice of $V_{\rm max}$ in Section~\ref{subsubsec:result_resolution}.

\subsubsection{Aspect Ratio and Domain Size}\label{subsubsec:result_aspect_ratio}

Some of the simulations listed in Table~\ref{tab:HSE_summary_params} are performed with different aspect ratios. While the fiducial box size is $L_{x}\times L_{y}=400\times50$, we have some simulations with longer $L_{x}$ or shorter $L_{y}$. First we discuss the effect of varying $L_{x}$, then the effect of varying $L_{y}$.

We increase $L_{x}$ in streaming dominated runs with larger CR flux, because the larger acceleration requires large box size to capture the gas - CR interactions. For the higher CR flux runs, the efficiency of momentum and energy transfer between CR and gas depends on $L_{x}$.  We define the $x$ direction momentum transfer efficiency $\dot{p}_{\rm {\rightarrow g}}/\dot{p}_{x, \rm CR}$ and energy transfer efficiency $\dot{E}_{\rm{\rightarrow g}}/\dot{E}_{\rm CR}$ as:

\begin{eqnarray}
    &\dot{p}_{x,\rm{\rightarrow g}}/\dot{p}_{x,\rm CR}=\frac{\int \sigma_{\rm c}\hat{\mathbf{n}}_{x}\cdot[\Fc-\vel\cdot(E_{\rm c}\textsf{I}+\textsf{P}_{\rm c})] dV}{\int_{\rm in}\mathcal{F}_{F_{\rm c}}dA} \label{eq:momentumeff}\\
    &\dot{E}_{\rm{\rightarrow g}}/\dot{E}_{\rm CR}=\frac{\int(\vel+\vs)\cdot(\nabla\cdot\textsf{P}_{\rm c})dV}{\int_{\rm in}\mathcal{F}_{E_{\rm CR}}dA}\label{eq:energyeff}
\end{eqnarray}
where $\mathcal{F}_{F_{\rm c}}$ and $\mathcal{F}_{E_{\rm CR}}$ are the flux of CR momentum in $x$ direction and CR energy returned by Riemann solver. So the $x$ direction momentum (or energy) transfer efficiency describes what fraction of total injected $x$ direction CR momentum (or CR energy) is given to the gas in the domain.

\begin{figure}
    \centering
    \includegraphics[width=\linewidth]{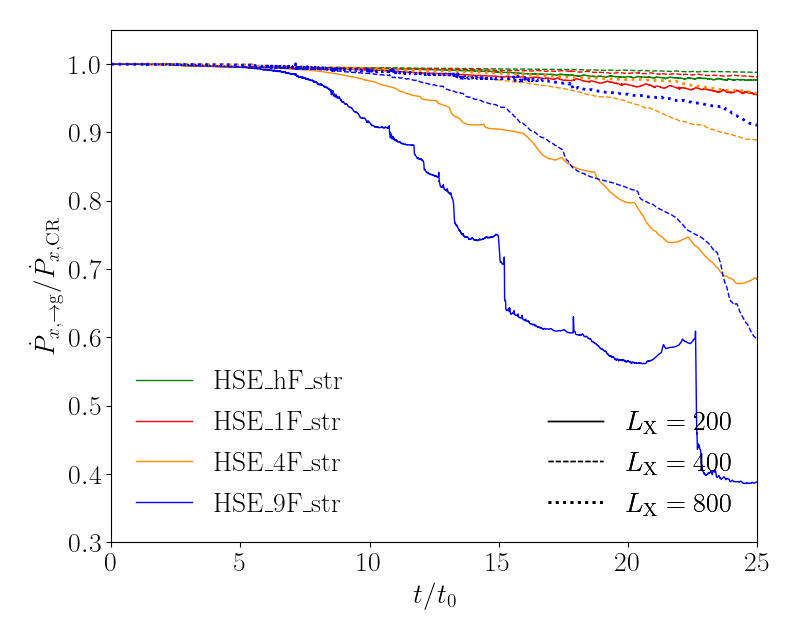}
    \caption{The effect of box size on momentum transfer efficiency. The lines are time evolution of the ratio between total momentum transferred to gas from CR and total injected CR momentum for HSE\_hF\_str (green lines), HSE\_1F\_str (red lines), HSE\_4F\_str (orange lines) and HSE\_9F\_str (blue lines). The line style shows the domain width $L_{x}$, where solid lines are for $L_{x}=200$, dashed lines are for $L_{x}=400$ (fiducial), and dotted lines are for $L_{x}=800$. }
    \label{fig:momratio_LX}
\end{figure}

Figure~\ref{fig:momratio_LX} shows the momentum transfer efficiency a function of time for different choices of box length $L_{x}$. In general, $L_{x}$ has a relatively insignificant effect on the simulations with sub-Eddington fluxes (red and green curves). However, the box size largely impacts the result of super-Eddington cases where the gas outflow is significant. For example in HSE\_9F\_str (blue curves), both momentum and energy transfer efficiency between CR and gas increases significantly as we go from shorter ($L_{x}=200$) to longer ($L_{x}=800$) boxes due to the capturing of more out-moving gas. Equivalently, the vertically longer domain allows more time for the CR forces to act on the gas, transferring a larger momentum fraction.

\begin{figure}
    \centering
    \includegraphics[width=\linewidth]{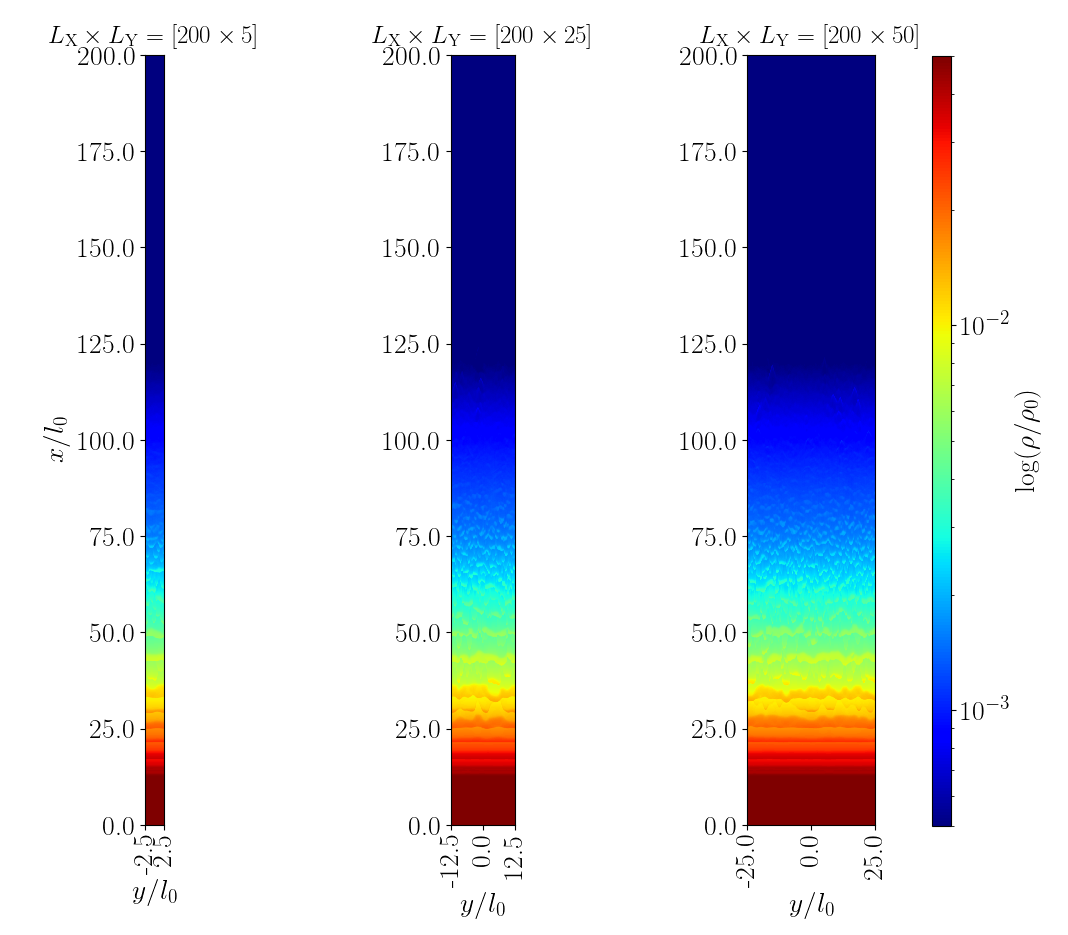}
    \caption{The gas density snapshots from HSE\_1F\_cd at $t=10.0$. From left to right: $L_{y}=5,25,50$. The aspect ratio of the third row is adjusted for the purpose of visual comparison, the $y/l_{0}$ labels denote the actual width.}
    \label{fig:densityLY}
\end{figure}

For our nearly plane-parallel geometry and nearly vertical magnetic field, the width of the simulation domain $L_{y}$ seems to have negligible results on the overall dynamics.  Figure~\ref{fig:densityLY} shows the gas density snapshots at $t=10.0$ for HSE\_1F\_cd, in which both streaming and diffusion are operating with $\kdiff=10.0$. There is a modest impact on the growth of the acoustic instability, but the differences have insignificant impact on the dynamics in the vertical direction.

\subsubsection{Resolution, Dimensionality and $V_{\rm max}$}\label{subsubsec:result_resolution}

\begin{figure}
    \centering
    \includegraphics[width=\linewidth]{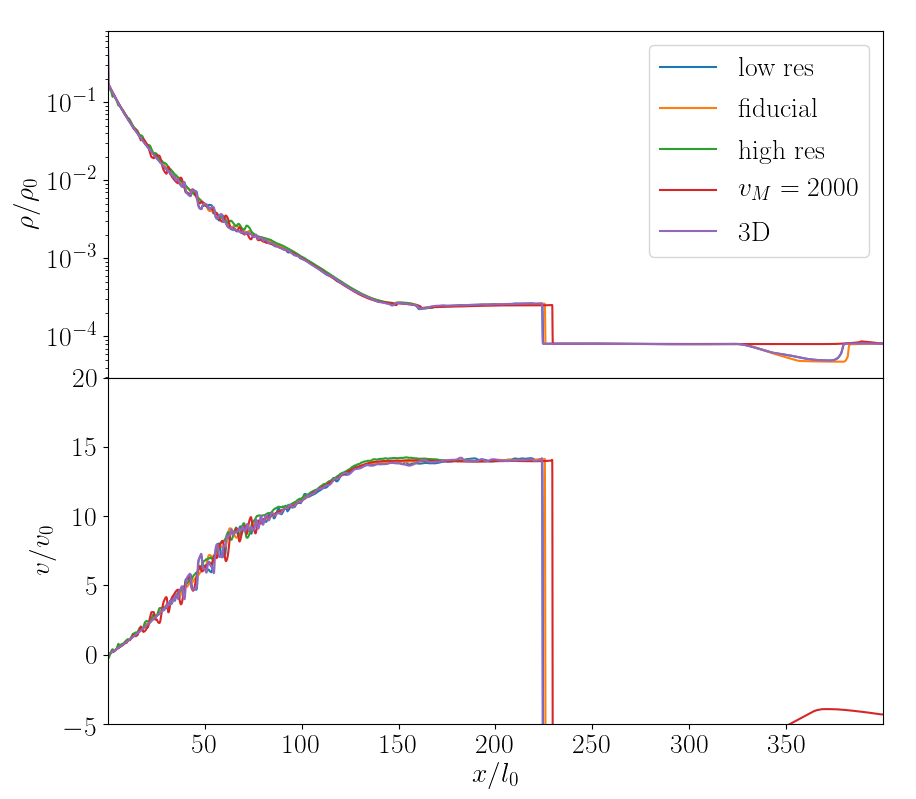}
    \caption{The $y$ direction averaged gas density (upper panel) and $x$ direction velocity (lower panel) of HSE\_1F\_str\_LR (blue), HSE\_1F\_str (orange), HSE\_1F\_str\_HR (green), HSE\_1F\_str\_VM (red). HSE\_1F\_str\_3D (purple) is the 3D simulation that is performed with the same resolution as HSE\_1F\_str\_LR, in which we show the profile averaged through $y$ and $z$ directions.}
    \label{fig:2dplotResolution}
\end{figure}

Figure~\ref{fig:2dplotResolution} shows a comparison of results at $t =10$ from high resolution (HSE\_1F\_str\_HR), moderate resolution (HSE\_1F\_str), and low resolution (HSE\_1F\_str\_LR) runs.  We fix the vertical and horizontal dimensions, increasing the number of zones in both directions by a factor of two for each increased resolution run. Aside from a modest effect on the development of the acoustic instability, the runs are insensitive to the resolution adopted here. The momentum and energy transfer efficiencies between gas and CR are also insensitive to resolution. Note that the amplitude of the shocks depends on resolution, with lower amplitude in the HSE\_1F\_str\_LR run, and nearly disappearing in even lower resolution test runs (not shown). But, the amplitude seems to be converging between HSE\_1F\_str\_HR and HSE\_1F\_str.

To study the impact of our two-dimensional assumption, we also perform a three-dimensional simulation, which is shown as the purple curve in Figure~\ref{fig:2dplotResolution}.  As with resolution, there is no evidence of significant dependence on dimensionality, at least for the streaming simulations and magnetic field geometry employed here.

Finally, we consider the dependence on $V_m$, which acts like a reduced speed of light for the time-dependent term in Equation~\ref{eq:crmomentum}.  This approximation allows us to take larger timesteps, which scale inversely with $V_{\rm m}$.  Previous tests have indicated that this is a reasonable approximation as long as $V_{\rm m}$ is larger than other physical velocities in the simulation. In simulations presented here, the maximum physical speed tends to be the Alfv\'{e}n speed in the diffuse background medium, which is about $v_{\rm A,max}\approx224$, while we use $V_{\rm m}=1000$ in all the simulations except HSE\_1F\_str\_VM, where $V_{\rm m}=2000$. In our dynamical problems, the time-dependent term $(1/V_{\rm m}^{2})\partial \Fc/\partial t$ in Equation~\ref{eq:crmomentum} is much smaller than the other terms, so we expect our choice of $V_{\rm m}$ should not impact our results. \citet{2018ApJ...854....5J} point out that the main physical effect of using $V_{\rm m}$ is overestimation of the momentum carried by CRs by $c^{2}/V_{\rm m}^{2}$, but this has little impact on the CR force, which determines momentum transfer to the gas.

The red curve in Figure~\ref{fig:2dplotResolution} shows the density snapshot of HSE\_1F\_str\_VM, also at $t=10.0$. The gas distribution and velocity are similar to others, with only modest differences in the development of the acoustic instability. The momentum and energy transfer efficiency agree with the $V_{\rm m}=1000$ simulations.

\section{Discussion}\label{sec:discussion}

\subsection{Momentum and Energy Transfer}

Our simulation results described in section~\ref{sec:results} explore a number of factors that impact the momentum and energy exchange between CRs and gas. To further quantify this exchange, we define the momentum efficiency $\dot{p}_{\rm x, to gas}/\dot{p}_{\rm x,CR}$ and energy efficiency $\dot{E}_{\rm to gas}/\dot{E}_{\rm CR}$ in Equation~\ref{eq:momentumeff} and Equation~\ref{eq:energyeff} respectively by taking the ratio of total momentum (energy) transferred from CRs to gas and the total injected CR momentum (energy). Table~\ref{tab:exchange_diff} tabulates the time averaged efficiencies at a point when these simulations roughly reach quasi-steady state.

\begin{table*}
\caption{Energy and Momentum Budget\tnote{a}}
\label{tab:exchange_diff}
\centering 
\begin{threeparttable}
\begin{tabular}{lcccc}
\hline
Name & $\dot{p}_{x,\rm{\rightarrow g}}/\dot{p}_{\rm x,CR}$  & $\dot{E}_{\rm{\rightarrow g}}/\dot{E}_{\rm CR}$ & $\dot{E}_{\rm{\rightarrow g,KE}}/\dot{E}_{\rm{\rightarrow g}}$ & $\dot{E}_{\rm{\rightarrow g,H}}/\dot{E}_{\rm{\rightarrow g}}$\\
\hline
HSE\_hF\_str & $98.9\%$ & $65.0\%$ & $4.1\%$ & $95.9\%$\\
HSE\_1F\_str & $98.4\%$ & $59.2\%$ & $7.1\%$ & $92.9\%$\\
HSE\_4F\_str & $96.9\%$ & $52.0\%$ & $16.7\%$ & $82.3\%$\\
HSE\_9F\_str & $93.8\%$ & $44.2\%$ & $24.1\%$ & $74.9\%$\\ 
HSE\_20F\_str & $77.3\%$ & $29.1\%$ & $35.1\%$ & $64.9\%$\\ 
HSE\_1F\_ld & $96.2\%$ & $52.9\%$ & $6.9\%$ & $93.1\%$\\
HSE\_1F\_cd & $96.7\%$ & $53.7\%$ & $6.4\%$ & $93.6\%$ \\
HSE\_1F\_hd & $96.7\%$ & $52.7\%$ & $8.3\%$ & $91.7\%$\\
HSE\_1F\_hd\_aniso & $96.8\%$  & $52.6\%$ & $8.3\%$ & $91.7\%$\\
HSE\_1F\_hd\_ns & $99.9\%$ & $27.9\%$ &  $100.00\%$ & -\tnote{b}\\
HSE\_1F\_ld\_ns & $99.9\%$ & $18.6\%$ &  $100.00\%$ & -\tnote{b} \\
HSE\_1F\_str\_b1 & $99.0\%$ & $56.8\%$ &  $16.0\%$ & $84.0\%$\\ %weak b field
HSE\_1F\_str\_b4 & $96.5\%$ & $53.0\%$ &  $4.0\%$ & $96.0\%$\\ %strong b field
\hline
\end{tabular}
\begin{tablenotes}
    \item[a] The energy and momentum exchange efficiency is averaged over $t=20.0t_{0}-25.0t_{0}$. The time interval is as when the simulation reaches quasi-steady exchange rate.  The simulations in this table correspond to the $L_{y}$ reported in Table~\ref{tab:HSE_summary_params} to avoid significant CR momentum and energy outflow.
    \item[b] Because streaming absent, there is no direct CR heating.
\end{tablenotes}
\end{threeparttable}
\end{table*}

The first column of Table~\ref{tab:exchange_diff} lists the momentum exchange efficiencies. In most of the simulations, $\gtrsim 95\%$ of momentum carried by the injected CR flux is transferred to gas within the simulation box, whether the flux is transported primarily by streaming or diffusion. The exception is HSE\_20F\_str, in which the momentum efficiency is $\sim 77\%$. The relatively low efficiency is due to rapid acceleration and subsequent outflow produced in this simulation. Figure~\ref{fig:momratio_LX} shows that small boxes cannot adequately capture the acceleration region, so the expectation is that the CR flux escaping through the upper boundary would continue to accelerate the gas if the simulation domain were extended.

Other effects that would limit coupling, such as preferential escape along low impedance channels appears to be absent.  This contrasts with radiatively accelerated outflows, where the impact of the radiative Rayleigh-Taylor instability likely limits the coupling between radiation and dusty gas by creating low density channels through with radiation can preferentially escape \citep{2012ApJ...760..155K,2014ApJ...796..107D}.  It is not entirely clear whether similar effects would be expected to be present in CR driven outflows, but the non-linear development of instabilities such as the CR acoustic instability \citep{1994ApJ...431..689B} or Parker instability \citep{1966ApJ...145..811P,2003ApJ...589..338R,2018ApJ...860...97H} could plausibly act in a fashion similar to Rayleigh-Taylor instability in radiatively driven flows. Indeed, we attribute the shocks seen in our simulations to the CR acoustic instability, as discussed further in section~\ref{subsec:discussion_ins} below.  In these  simulations, however, the growth of the instability does not seem to be modifying the momentum coupling, although it concentrates heating near the pressure jumps.  Nor do we see any evidence of the Parker instability, although this is likely impacted by our adoption of a nearly vertical field in the acceleration region.  Other simulation results suggest this instability may play an important role in overall structure of the gas disk \citep{2020ApJ...891..157H}.

The transfer of energy between the CRs and gas is more complex. The third column in Table~\ref{tab:exchange_diff} lists the time-averaged total energy transfer efficiency, and the fourth and fifth column list the fractions of kinetic energy and CR heating respectively. As with momentum transfer, we find that the CR energy transfer is sensitive to our domain size, with outflow from the top boundary carrying away a significant CR energy flux that is expected to generate further heating in larger boxes. Hence, it is difficult to judge the overall level of coupling, which will inevitably depend on the global structure of the flow and galaxy potential.  Nevertheless, we can draw conclusions about the relative importance of CR heating to CR acceleration.

Comparison of the third and fourth columns shows that for streaming runs, the majority of the energy transfer is due to CR heating. This is true even in hybrid simulations where diffusion is significant. In contrast, there is no explicit CR heating in the diffusion runs, due to the assumptions of our CR transport model and all energy transfer increases the kinetic energy of the gas.

\begin{figure}
    \centering
    \includegraphics[width=\linewidth]{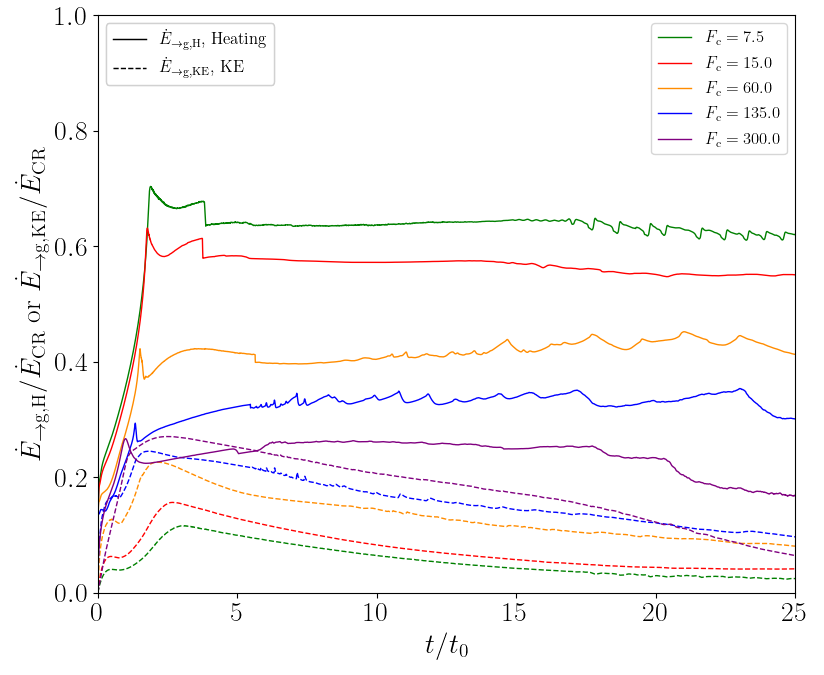}
    \caption{The energy transfer efficiency as a function of time for simulations with diffusion. The total energy transferred to the gas from CR includes heating (the solid line) and kinetic energy (the dashed line). HSE\_1F\_str (red) is the streaming simulation with $\kdiff=10^{-8}$. HSE\_1F\_ld (black), HSE\_1F\_cd (light blue) and HSE\_1F\_hd (green) are streaming-diffusion simulations with $\kdiff=1.0,~10.0,~100.0$ respectively. HSE\_1F\_ld\_str (brown) and HSE\_1F\_str (orange) are pure diffusion simulations with $\kdiff=1.0,~100.0$ respectively. }
    \label{fig:energyratio_diff}
\end{figure}

We show the time evolution of energy transfer components in  Figure~\ref{fig:energyratio_diff} for varying CR transport mechanisms. After an initial transient stage on the order of the Alfv\'{e}n wave crossing time, the efficiency is relatively constant. The two non-streaming simulations (brown and yellow dashed curves) have only a kinetic energy term, and share similar transfer efficiency corresponding to about $20\%-30\%$ of total injected CR energy. The CR heating is weakly sensitive to the diffusivity, showing a trend towards increased heating efficiency when CR diffusivity is lowered. 

\begin{figure}
    \centering
    \includegraphics[width=\linewidth]{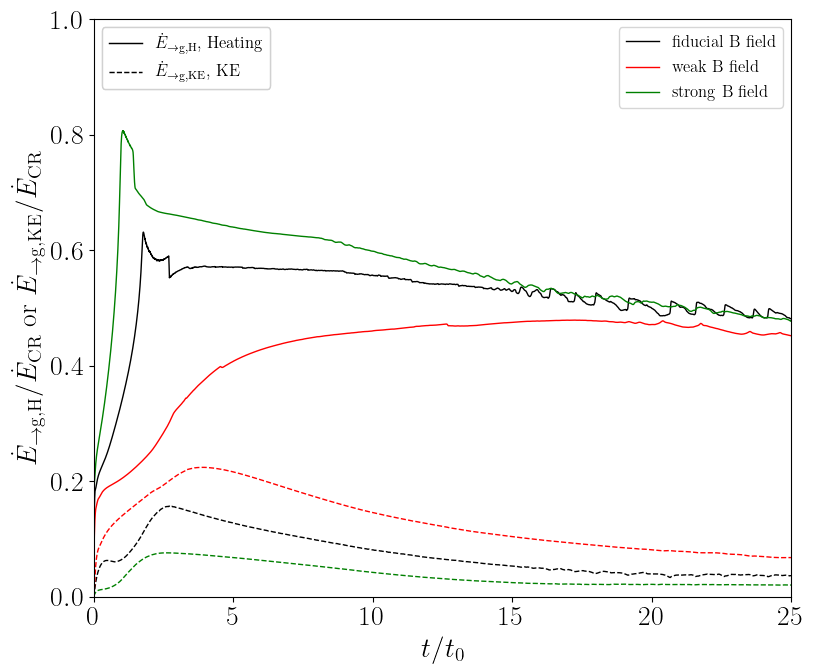}
    \caption{Similar to Figure~\ref{fig:energyratio_flux}, HSE\_1F\_str\_b1 (the red lines) has initial magnetic field $B'=1.0$ and HSE\_1F\_str\_b4 (the green lines) has initial magnetic field $B'=4.0$}
    \label{fig:energyratio_bfield}
\end{figure}

In Figure~\ref{fig:energyratio_bfield} we show how the energy transfer depends on the magnetic field strength by comparing runs with initial magnetic fields a factor of two larger (HSE\_1F\_str\_b4) or smaller (HSE\_1F\_str\_b1) than our fiducial run. For the range explored here, modifying the magnetic field modifies the CR heating's fractional contribution, but the results are relatively mild. The stronger (weaker) ${\bf B}$ in HSE\_1F\_str\_b4 (HSE\_1F\_str\_b1) lead to factor of two changes in the initial $v_{\rm A}$ relative to the fiducial run so one might expect commensurate changes in the heating rate, which is nominally proportional to $v_{\rm A}$ in our streaming setup.  This is qualitatively consistent with early times where the heating rates increase with increased magnetic field strength, but the systems assymptote to roughly the same values at late times. Even at early times, however, the increase (decrease) in $v_{\rm A}$ is offset by a decrease (increase) in $P_{\rm c}$ since our boundary condition targets a fixed $F_{\rm c}$.  At later times, the gradient of $P_{\rm c}$ is shallower for the higher magnetic field cases. The upshot is that dissipation varies with ${\bf B}$, but the dependence is somewhat weak over the range explored here for the constant $F_{\rm c}$ boundary condition.

\begin{figure}
    \centering
    \includegraphics[width=\linewidth]{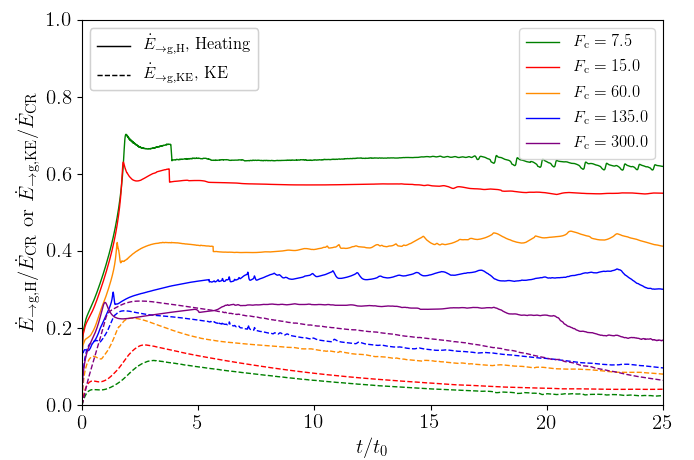}
    \caption{The energy transfer efficiency as a function of time for simulations with diffusion. The total energy transferred to the gas from CR includes heating (the solid line) and kinetic energy (the dashed line). These are all streaming simulations with different CR flux: HSE\_hF\_str (green), HSE\_1F\_str (red), HSE\_4F\_str (orange) and HSE\_9F\_str (blue) and HSE\_20F\_str (purple).}
    \label{fig:energyratio_flux}
\end{figure}

When streaming dominates the energy transfer, the overall efficiency and partition between kinetic energy and heating depends on the injecting CR flux as shown in Figure~\ref{fig:energyratio_flux}. A higher injected CR flux leads to smaller ratio between heating and kinetic energy. The kinetic energy term is proportional to fluid velocity $v_{\rm x}$, while the heating term is proportional to local Alfv\'{e}n velocity. The gas is accelerated to higher velocity in the larger CR flux simulations, so the kinetic energy term rises. In our highest CR flux run (HSE\_20F\_str), the kinetic energy and heating are roughly comparable to each other.

The end result is that CR acceleration of outflows is most efficient when heating is limited.  This is trivially achieved in our non-streaming runs where heating is absent.  This result is in qualitative agreement with the results found by \citet{quataert2021stream}, who found that CRs drive much weaker mass outflows when streaming is present than in the pure diffusion limit \citep{quataert2021physics}. Quantitatively, they see a larger difference between pure diffusion and streaming runs than we do, but this may be attributed to their use of an isothermal equation of state, which would be equivalent to adopting a much stronger cooling regime.  If streaming operates, increasing the injected CR flux accelerates the gas more rapidly, lowering the ratio of gas velocity to the streaming (Alfv\'{e}n) velocity. This highlights the importance of understanding the detailed dynamics of CR transport in star-forming galaxies and, if streaming is important, introduces stronger dependence on CR flux than might be anticipated by a simple comparison of the CR flux to our derived Eddington flux.

\subsection{Implications for CR Feedback in Galaxies}\label{subsec:discussion_galaxies}

In this section, we vary our characteristic scales to infer properties of CR outflows in different galactic environments. Our target location within a galaxy is the atmosphere of the gas disk, where a wind might be launched. We focus on winds launched solely by CRs, but of course other mechanisms such as supernova driving and radiation pressure are likely operating in real systems. This motivates characteristic densities and lengthscales intermediate between the disk interior and the extended halo.

Following  \citet{socrates2008eddington}, we estimate the CR flux $F_{\rm c}$ from galaxies based on the surface star formation rate $\dot{\Sigma}_{\star}$ and scaling the CR production as a fraction of the expected supernova rate.  Here we adopt the scaling of \citet{crocker2021cosmic}
\begin{equation}
    F_{\rm c} \simeq 1.85\times10^{-3}\left(\frac{\dot{\Sigma}_{\star}}{M_{\odot}~\rm kpc^{-2}\rm yr^{-1}}\right) \rm erg~cm^{-2}~s^{-1} \label{eq:fcr_galaxy}.
\end{equation}
This assumes a \citet{chabrier2005initial} initial mass function so that, on average, there is a supernova every $90M_{\odot}$ stellar mass. Each supernova event roughly converts $\sim10\%$ of kinetic energy $E_{\rm SN} \simeq 10^{51} \rm erg$ into CR energy. Assuming these assumptions are correct, this flux is an upper limit because it ignores any CR losses that may present, such as those resulting from pion production after collision with ISM particles or 'adiabatic' losses during transport \citep{socrates2008eddington,lacki2010,chanetal2019}.

In Section~\ref{subsec:result_eddingtonflux}, we derived an estimate for the CR Eddington flux by balancing the CR force with gravity. We estimate the gravity $g$ in Equation~\ref{eq:feddstreaming} according to disk surface density as $g=2\pi G\Sigma_{\rm gas}/f_{\rm gas}$, where $f_{\rm gas}$ is gas fraction, which we scale to 0.5 for simplicity. With these assumptions, the scaling between different galaxies is primarily dependent on choosing $\dot{\Sigma}_{\star}$ and $\Sigma_{\rm gas}$.

The remaining ambiguity is in the choice of $\rho$ to use in this estimate since it can vary significantly within the galaxy. The equations imply that for sufficiently low densities, there will always be some level of outflow, which is broadly consistent with our numerical results.  For small enough densities, however, we do not expect significant mass loss.  For simplicity we adopt $\rho = \Sigma_{\rm gas}/(2 z_{\rm d})$, where $z_{\rm d}$ is an estimate of disk scale height. We emphasize that this is an estimate that would be expected to yield significant feedback from the disk galaxy and not simply one that would drive an observable outflow.  Finally, for the streaming case we estimate $v_{\rm s} = \sqrt{2 k_B T/(\mu m_p \beta)}$, where $\beta$ is the ratio of magnetic to gas pressure. With these assumptions, the corresponding Eddington flux in streaming and diffusion limits are
\begin{eqnarray}
    F_{\rm{edd,str}} = 1.8\times10^{-4}\left(\frac{\Sigma_{\rm{gas}}}{100 M_{\odot}\rm pc^{-2}}\right)^{2}\left(\frac{T}{10^{4}K}\right)^{1/2}\nonumber \\
    \left(\frac{f_{\rm{gas}}}{0.5}\right)^{-1}\left(\frac{\beta}{1.0}\right)^{-1/2}
    \left(\frac{H_\rho}{0.1 z_{\rm d}}\right){\rm erg~cm^{-2}~s^{-1}},
\label{eq:fedd_str_sc}
\end{eqnarray}
and
\begin{eqnarray}
    F_{\rm{edd,diff}} = 5.9\times10^{-2}\left(\frac{\Sigma_{\rm{gas}}}{100 M_{\odot}\rm pc^{-2}}\right)^{2}\left(\frac{\kappa}{10^{29}\rm{cm^{2}s^{-1}}}\right)\nonumber\\
    \left(\frac{f_{\rm{gas}}}{0.5}\right)^{-1}\left(\frac{z_{\rm disk}}{100\rm{pc}}\right)^{-1}\rm{erg~cm^{-2}~s^{-1}}.
    \label{eq:fedd_diff_sc}
\end{eqnarray}

\begin{figure}
    \centering
    \includegraphics[width=\linewidth]{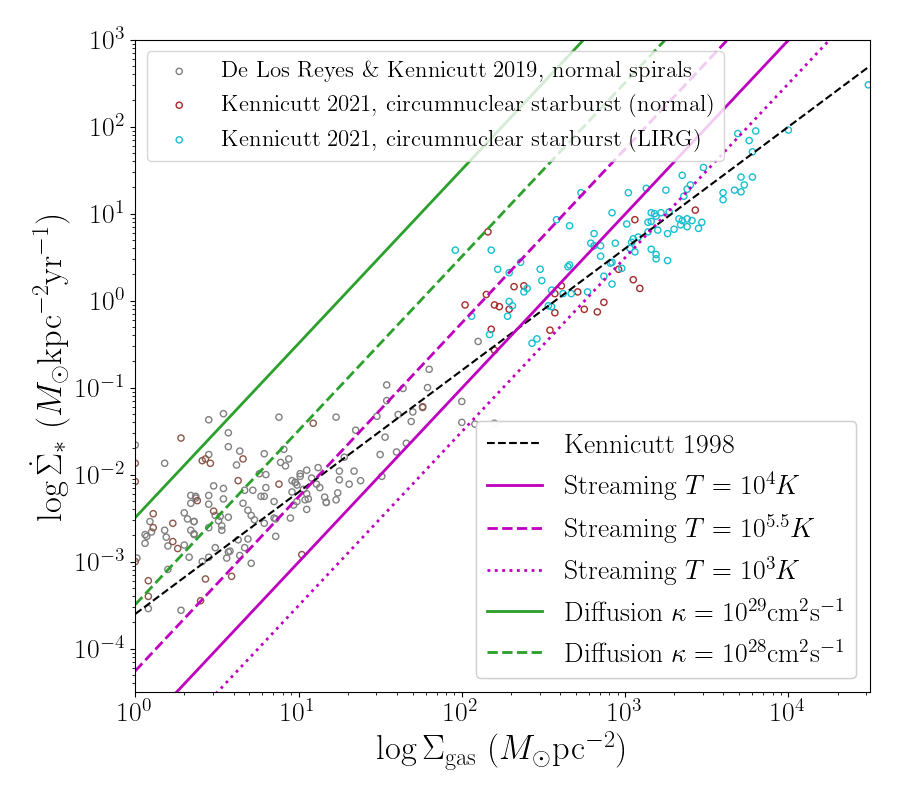}
    \caption{The magenta lines show the Eddington limit where the CR flux (Equation~\ref{eq:fcr_galaxy}) equals to CR Eddington flux in streaming limit (Equation~\ref{eq:fedd_str_sc}). The line style corresponds to different choices of temperature in Equation~\ref{eq:fedd_str_sc} (solid: $T=10^{4}$K, dashed:$T=10^{5.5}$K, dotted:$T=10^{3}$K). The green lines are in diffusion limit. The line style corresponds to different diffusivity in  Equation~\ref{eq:fedd_diff_sc} (solid: $\kdiff=10^{29}\rm{cm^{2}/s}$, dashed:$\kdiff=10^{28}\rm{cm^{2}/s}$). The black dashed line shows the classical Kennicutt-Schmidt relation \citep{kennicutt1998global}. The non-filled circles are observational data from \citet{de2019revisiting} and \citet{kennicutt2021revisiting}.}
    \label{fig:massloading_asmp_Sigma}
\end{figure}

We can estimate the surface density and star formation rates where a galaxy might be expected to exceed the CR Eddington limit and drive outflow by setting Equation~\ref{eq:fcr_galaxy} equal to either Equation~\ref{eq:fedd_str_sc} or Equation~\ref{eq:fedd_diff_sc} when streaming or diffusion (respectively) dominates. The results are shown as magenta and green solid lines in the upper panel of Figure~\ref{fig:massloading_asmp_Sigma}. The upper left region of each line is where the CR flux is above the Eddington limit, and vice versa. We also show normal galaxies measured in \citet{kennicutt2021revisiting} and \citet{de2019revisiting} in Figure~\ref{fig:massloading_asmp_Sigma}. 

We see that at low surface densities galaxies mostly lie near or above these estimates for the Eddington limit, suggesting that CR can drive powerful winds in these galaxies.  This could be an indication that we have either overestimated the CR flux in the wind launching region due to CR losses during transport through the gas disk or underestimated the Eddington ratio.  Alternatively, it may mean that these galaxies are all driving outflows due to CRs but that the mass loading factors are not large enough to completely quench star formation. At higher surface densities, it would seem that CRs alone are less efficient at driving outflows powerful enough so that they would feed back significantly on star formation in the disk. Our simple estimates are broadly consistent with those of \citet{crocker2021b}, who include further $\Sigma_{\rm gas}$ dependence in $f_{\rm gas}$, the disk scale height (through the ISM velocity dispersion), and streaming velocity to sound speed ratio.

Thus far, we have not utilized our simulation results directly, other than to justify the adoption of an Eddington flux condition for the driving of outflows. The primary value of the simulations is that they give us an estimate of the mass outflow rate $\dot{M}_{\rm out}$ for a given $F_{\rm c}$ and $g$. As discussed in section~\ref{subsubsec:UnitScaling}, the non-radiative simulations are scaled to physical units by choosing a density scale $\rho_{0}$, length scale $l_{0}$, and temperature scale $T_0$. These, in turn, are constrained by matching our simulation results to galaxies using physical units via
\begin{eqnarray}\label{eq:g_Fcr}
    \frac{2\pi G\Sigma_{\rm{gas}}}{f_{\rm{gas}}}&=&\left(\frac{k_{b}T_{0}}{\mu l_{0}}\right)g',\\
    F_{\rm{c}}&=&\rho_{0}\left(\frac{k_{b}T_{0}}{\mu} \right)^{3/2}F_{\rm{c}}'.
\end{eqnarray}
Here, $g'$ and $F_{\rm{c}}'$ have primes to denote the dimensionless gravity and injected CR flux prescribed in the simulation, and the left hand side values are in physical units. Using Equation~\ref{eq:fcr_galaxy}, we can rewrite our characteristic scales as
\begin{eqnarray}
    &l_{0}=21\left(\frac{T_{0}}{10^4 K}\right)\left(\frac{\Sigma_{\rm{gas}}}{100 M_{\odot}\rm pc^{-2}}\right)^{-1}\left(\frac{f_{\rm gas}}{0.5}\right)\left(\frac{g'}{0.826}\right){\rm pc},\label{eq:l0}\\
    &\rho_{0}=7.6\times10^{-23}\left(\frac{F_{\rm c}'}{15}\right)^{-1}
    \left(\frac{T_{0}}{10^4 K}\right)^{-3/2}\left(\frac{\dot{\Sigma}_{\star}}{M_{\odot}/\rm kpc^{2}/\rm yr}\right) {\rm g/cm^{3}}.\label{eq:rho0}.
\end{eqnarray}
Once we fix the injecting CR flux, each data point in the $\Sigma_{\rm gas}-\dot{\Sigma}_{\star}$ plane provides two constraints on the scaling parameters. We adopt the convention of choosing the temperature and using $\Sigma_{\rm gas}$ and $\dot{\Sigma}_{\star}$ to determine $l_0$ and $\rho_0$ for given $T_0$.

With these scaling we can estimate the column of material impacted by our CR driven outflows via
\begin{eqnarray}
\frac{\rho_0 l_0}{\Sigma_{\rm gas}} = 0.24 \left(\frac{F_{\rm c}'}{15}\right)^{-1}
\left(\frac{T_{0}}{10^4 K}\right)^{-1/2}\left(\frac{\dot{\Sigma}_{\star}}{M_{\odot}/\rm pc^{2}/\rm Myr}\right)\nonumber\\
\left(\frac{\Sigma_{\rm{gas}}}{100 M_{\odot}\rm pc^{-2}}\right)^{-2}\left(\frac{f_{\rm gas}}{0.5}\right)\left(\frac{g'}{0.826}\right).
\end{eqnarray}
For objects lying on or near the Kennicutt-Schmidt relation, this will lead to a decrease in the fraction of the atmosphere impacted at higher $\Sigma_{\rm gas}$. We can also estimate the mass outflow per unit area as $\dot{M}_{\rm{out}}=\dot{m}'v_0 \rho_0$, where $\dot{m}'$ is the dimensionless outflow mass per area in our simulations. We find
\begin{eqnarray}
\frac{\dot{M}_{\rm out }}{\dot{\Sigma}_{\star}} = 0.034 \left(\frac{F_{\rm c}'}{15}\right)^{-1}\left(\frac{\dot{m}'}{0.00252}\right)
\left(\frac{T_{0}}{10^4 K}\right)^{-1/2}.
\end{eqnarray}
Our simulations yield $\dot{m}' = 0.00252$, 0.0130, 0.0396, and 0.0707 for $F_{\rm c}' = 15$, 60, 135, and 300, respectively or approximately $\dot{m}' \propto F_{\rm c}'$ over the range we have explored.  This means that the implied mass loading factor is approximately constant for the simulations and is a few percent of the star formation rate for $T_0 \simeq 10^4$K. It is important to emphasise that the density at the base of the simulation has a large effect on the implied mass loading in all models. Equation~\ref{eq:rho0} shows that for a given $\dot{\Sigma}_{\rm \star}$, the highest densities correspond to the lowest temperatures, with the highest mass loading corresponding to $T_0 \lesssim 10^3$K. 

A lower temperature is broadly consistent with the larger fractions of dense molecular gas in LIRGS and ULIRGs but poses a self-consistency problem for the CR transport models adopted here, which assume the gas is sufficiently ionised so that CRs stream at nearly the Alfv\'{e}n speed and have diffusivities similar to the Milky Way.  If ionisation fractions are low, the effective Alfv\'{e}n speed may be significantly higher. Furthermore, the streaming instability or turbulent fluctuations that underlie diffusive models may be strongly damped in molecular gas \citep{bustard2020cosmicraytransport}. This may increase the escape of CRs from the highest density regions and reduce the impact of destruction processes like pion production, but could also lower the coupling between CRs and outflowing gas unless sufficiently high ionisation fractions are present in the wind launching regions of the galaxy, which are presumably closer to the disk surface.  Resolving these questions is beyond the scope of this work, but is central to addressing the question of how much CR feedback can impact galaxies with high molecular gas fractions.

\subsection{Acoustic Instability}\label{subsec:discussion_ins}
In all our simulations with streaming, a series of shocks grow and saturate while gas is being accelerated by CRs. For example, the shocks in the gas density and velocity in Figure~\ref{fig:1Fstrlineplot}, or the horizontal bands in the 2D gas density snapshots in Figure~\ref{fig:density_flux} and Figure~\ref{fig:density_diff}. These shocks do not form in our diffusive only simulations. We find that the growth of these shocks are consistent with cosmic rays acoustic instability, and our present whenever the resolution is sufficiently high.

\citet{1994ApJ...431..689B} studied the CR acoustic instability in the streaming limit with a uniform background. They found that the growth rate varies with $m\equiv v_{\rm A}/c_{\rm s}$ and $c\equiv c_{\rm CR}/c_{\rm s}$, where $v_{\rm A}$, $c_{\rm s}$ and $c_{\rm CR}$ are Alfv\'{e}n speed, sound speed and CR acoustic speed (see definitions in Appendix~\ref{sec:appendixA}). \citet{tsung2021cosmic} extends this work and discusses the physical origin of the shock structures and various relevant microphysical processes. In this section, we refer to the gradient in gas density, gas pressure and CR pressure as the ``background gradient'', and derive the dispersion relation in Appendix~\ref{sec:appendixA}.

The emergence and growth of shocks in our simulation is spontaneous and dynamical.  They usually appear after the gas density and pressure are redistributed to the larger scale height compared to the initial condition. The shocks are approximately even-spaced and have similar amplitude. We measure the average shock amplitude increment over time to estimate the acoustic instability growth rate.

\begin{figure}
    \centering
    \includegraphics[width=0.45\textwidth]{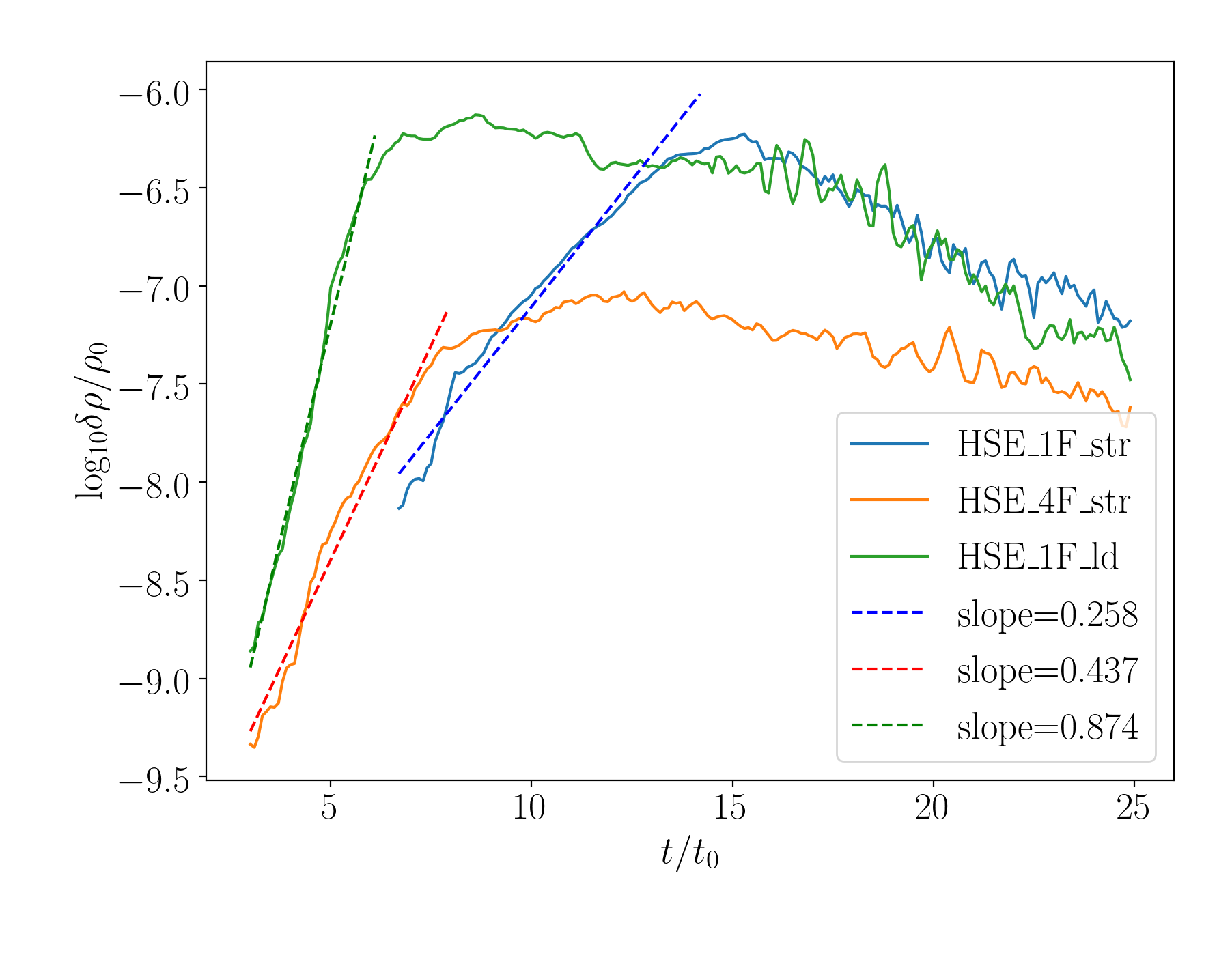}
    \caption{The growth rate of the shocks caused by acoustic instability for HSE\_1F\_str (blue), HSE\_4F\_str (orange) and HSE\_1F\_ld (green). The dashed line is the fitted growth rate, with slopes noted in the legend. The estimated slopes from solving the dispersion relation are 0.2, 0.5 and 0.8 respectively.}
    \label{fig:Diss_growth}
\end{figure}

We first filter the high spatial frequency shocks and fit a smooth curve to the $y$ direction averaged gas density using the LOWESS (Locally Weighted Scatterplot Smoothing) method. Then we take the weighted volume average residual as the shock amplitude $\delta\rho'$. We show the resulting amplitude evolution as the solid curves in Figure~\ref{fig:Diss_growth} for three simulations. The shocks experience an initial linear growth, which we fit with the dashed lines, before dropping at a later time. The drop in the average amplitude at later time reflects that the shocks begin to be advected out of the domain with the fluid.

We solve dispersion relation of the CR acoustic instability (see Appendix~\ref{sec:appendixA}) to obtain theoretical growth rates and compare them with the above measurements. Our calculation is based on the sampling relevant $y$ direction averaged quantities at the time and location where we first see the shocks. The calculation for HSE\_1F\_str is based on the profiles at $t=6.7t_{0}$ and $x=33.0l_{0}$, HSE\_4F\_str is based on the profiles at $t=3.6t_{0}$ and $x=25.0l_{0}$, and HSE\_1F\_ld uses profile at $t=3.2t_{0}$ and $x=25.0l_{0}$. The estimated growth rates are $0.2$, $0.5$ and $0.8$ for HSE\_1F\_str, HSE\_4F\_str and HSE\_1F\_ld respectively (the corresponding slopes in Figure~\ref{fig:Diss_growth}). The good correspondence between estimated growth rates and measured growth rates is consistent with shocks being seeded by the acoustic instability. 

The expected growth rates in our simulation are relatively low due to the low ratio between CR pressure and gas pressure in the diffuse region.  Figure~\ref{fig:Diss_ins_weakdiffstreaming} shows the growth rate for the $k=1$ mode, the yellow circles label some typical $c_{\rm CR}/\cs-v_{\rm A}/\cs$ pairs in our simulations near the position and time the shocks starts to grow. The ``bottleneck'' slows down CR propagation in dense gas and yields a low CR pressure in diffuse region, while the gas is heated by the initial CR shock. As a result, the low $c_{\rm CR}/c_{\rm cs}$ and high $v_{\rm A}/c_{\rm s}$ constrains the growth rate to low level in the simulations.

The instability eventually saturates and the shock amplitudes are relatively constant over time. The shocks spacing increase as the pressure scale heights increase \citep{tsung2021cosmic}, and advects with the gas. The instability creates multiple ``stairs'' in CR pressure. The steep pressure gradients at shock fronts dominate the CR heating at later time when the overall CR pressure gradient is shallow.

\section{Summary and Conclusions}\label{sec:conclusion}

Many studies have identified CR pressure and heating as promising mechanisms for driving galactic outflows and ISM turbulence. In this work, we studied the dynamical interaction due to an imposed CR flux at the base of an initially hydrostatic atmosphere to investigate the outflow launching process. We derived estimates for the CR Eddington flux in the streaming limit (equations~\ref{eq:feddstreaming}) and diffusion limit (equations~\ref{eq:fedddiff}). We performed simulations with different CR transport models including streaming, streaming-diffusion and diffusion. In the set of streaming simulations, we vary the injected CR flux from sub-Eddington to super-Eddington to quantify how this impacts the atmosphere and outflow that develops. We measured the outflow rate from simulations and scale simulated systems to realistic star forming environments. We also consider simulations with radiative cooling, showing that cooling modifies the wind density and pressure profile, resulting in generally more inhomogeneous and cooler outflow. 

We measured the momentum and energy transfer efficiency from CR to gas, and summarised how it changes with the injected CR flux and the relative importance between streaming and diffusion. We summarise our results as follows:

\begin{itemize}
  
\item Star forming systems near the classical Kennicutt-Schmidt relation are likely to produce super-Eddington CR fluxes (Figure~\ref{fig:massloading_asmp_Sigma}) if a large fraction of the CRs survive transfer through the gas to reach the surface.  This contribution should be most important in environments with lower star-formation rates and surface densities, where even relatively high gas densities near the midplane are impacted. More generically, we expect that for a wide range of galaxy properties, there will be a column of low density gas that can be driven to outflow, but this represents a smaller fraction of the gas in galaxies with higher surface densities. The expected mass outflow is a few percent of the star formation rate for an initially warm atmosphere ($\sim10^{4}K$).
    
\item We identify the growth and saturation of CR acoustic instability in our streaming simulations. However, the presence of instability does not limit the momentum transfer from CR to gas. The momentum transfer efficiency is usually $\gtrsim95\%$ regardless the CR flux and does not strongly depend on whether streaming or diffusion dominates transport, suggesting that CR and gas remain well-coupled in our simulations.
    
\item The energy transfer between CR and gas is sensitive to CR flux and the dominant CR transport mechanism. The fraction of total energy transferred to gas from CRs increases with CR flux. When streaming is included, a substantial fraction of transferred energy is in the form of Alfv\'{e}n wave heating, resulting in larger pressure scale height and outflowing gas hotter than the initial temperature by more than one order of magnitude. Typically $\lesssim10\%$ of the total CR energy is imparted to gas kinetic energy. Pure diffusion  without the Alfv\'{e}n wave heating leads to roughly twice as much energy being transferred to the kinetic energy of the outflow.
    
    \item For the transport scheme used here, streaming tends to limit the CR force applied to the gas for a given CR flux.  In the absence of streaming, a lower diffusivity requires a larger CR pressure gradient to transport the same CR flux, leading to a large CR force on the gas.  In contrast, when streaming operates with a low diffusivity and a relatively high streaming speed, it will dominate the transport of CRs relative to diffusion, providing a weaker CR pressure gradient and a lower CR force.  For a sufficiently high diffusivity, CR diffusion can again dominate the transport but will then produce a weaker CR pressure gradient and subsequently weaker force.
    
    \item Radiative cooling is potentially important, especially in sub-Eddington and near-Eddington systems in terms of setting the gas temperature and pressure profile of the extended atmosphere and outflow. When the CR flux is highly super-Eddington, CR pressure is more important than gas pressure and cooling has a weaker effect on mass loss. Cooling significantly lowers the wind temperature, promoting gas condensation near the density irregularities seeded by acoustic instability, resulting generally cooler and less homogeneous outflow.
    
\end{itemize}

\section*{Acknowledgements}
We would like to thank Yan-fei Jiang for sharing the cosmic rays module implemented in Athena++ as well as for support and throughout the project. We would also thank Loreto Barcos-Mu\~noz, S. Peng Oh, Yiqing Song, Stephanie Tonnesen, Tsun Hin Navin Tsung, and Dong Zhang for helpful discussions and insights. This work used the computational resources provided by the Advanced Research Computing Services (ARCS) at the University of Virginia. We also used the Extreme Science and Engineering Discovery Environment (XSEDE), which is supported by National Science Foundation (NSF) grant No. ACI-1053575. This work was supported by the NSF under grant AST-1616171. X. Huang was supported by the Simons Foundation through the Flatiron Institute’s Pre-Doctoral Research Fellowship program while a portion of this work was completed.

\section*{Data Availability}
The Athena++ code is publicly available at https://github.com/PrincetonUniversity/athena. The cosmic ray module used here is not yet part of the public release but was obtained from Dr. Yan-Fei Jiang who made it available upon request.  The initialization files, simulation outputs, and data analysis scripts for this study will be shared upon request to the corresponding author.

\bibliographystyle{mnras}
\bibliography{main}

\appendix 
\section{Acoustic Instability Dispersion Relation}\label{sec:appendixA}

We briefly summarise the acoustic instability dispersion relation based on CR transport equations for the conditions relevant to our simulations. We refer the reader to \citet{tsung2021cosmic} for a more thorough discussion of the instability . To keep the problem tractable, we assume a one-dimensional geometry with a 1-moment approximation similar to that used in previous work \citep{1994ApJ...431..689B}. The 1-moment and 2-moment scheme gives similar results in the derived growth rate \citep{tsung2021cosmic}. We first define the scale height for the background CR pressure $\Lc$, gas pressure $\Lg$ and gas density $L_{\rho}$ as:
\begin{eqnarray}\label{eq:discuss_scale_height}
    L_{\rm{c}}^{-1}\equiv-\frac{\partial\ln\pcr}{\partial x}, \quad
    L_{\rm{g}}^{-1}\equiv-\frac{\partial\ln\pg}{\partial x}, \quad
    L_{\rho}^{-1}\equiv-\frac{\partial\ln\rho}{\partial x}
\end{eqnarray}
We also define the scaled diffusivity $l_{\rm c}$
\begin{equation}
    l_{\rm c}\equiv\frac{\kdiff}{c_{\rm s}}
\end{equation}
Using WKB approximation with perturbations $\propto \exp(i\omega t - i k x)$, when $kL\gg1$, we can rewrite the perturbed equations as:
\begin{eqnarray}
    \omega\delta\rho&=&\bigg(1-\frac{i}{kL_{\rho}}\bigg)k\rho\delta v\nonumber\\
    \omega\delta v&=&\frac{k}{\rho}\delta\pg+\frac{k}{\rho}\delta\pcr-i\frac{\delta\rho}{\rho}g\nonumber\\
    \omega\delta\pg&=&\bigg(1-\frac{i}{\gamma_{\rm g}k\Lg}\bigg)\gamma_{\rm{g}}k\pg\delta v\\
    &+&i(\gamma_{\rm{g}}-1)\frac{\pcr v_{\rm{A}}}{\Lc}\frac{\delta\rho}{2\rho}+(\gamma_{\rm{g}}-1)k v_{\rm{A}}\delta\pcr\nonumber\\
    \omega\delta\pcr&=&\bigg(1-\frac{i}{\gamma_{\rm{c}}k\Lc}\bigg)\gamma_{\rm{c}}k\pcr\delta v \\
    &+&(kv_{\rm{A}}+i\kdiff k^{2}+\frac{i\gamma_{\rm{c}}v_{\rm A}}{2L_{\rho}})\delta\pcr \\
    &-&(1+\frac{3i}{2kL_{\rho}}-\frac{i}{\gamma_{\rm{c}}k\Lc})\gamma_{\rm{c}}kv_{\rm{A}}\pcr\frac{\delta\rho}{\rho}
\end{eqnarray}
The quantities with $\delta$ are the perturbed quantities, $\omega$ is the frequency and $k$ is the wave number. To compare with \citet{1994ApJ...431..689B} and other previous work, these equations differ slightly from the equations evolved by Athena++ in two ways.  Equation~(\ref{eq:crmomentum}) is replaced by a relation for the CR flux of the form
\begin{eqnarray}
F_{\rm c} = (v+v_{\rm A})(E_{\rm c} + P_{\rm c}) - \kdiff\frac{\partial E_{\rm c}}{\partial x}. 
\end{eqnarray}
This is generally a good approximation since the $\partial F_{\rm c}/\partial t$ term in equation~(\ref{eq:crmomentum}) is negligible. Furthermore, previous treatments replace $\partial E_{\rm c}/\partial t$ with $\partial P_{\rm c}/\partial t$ in equation~(\ref{eq:crenergy}).

We adopt \citet{1994ApJ...431..689B} notation and define:
\begin{equation}
    \nu\equiv\frac{\omega}{kc_{\rm{s}}},\quad m\equiv\frac{v_{\rm{A}}}{c_{\rm{s}}},\quad c\equiv\frac{c_{\rm CR}}{c_{\rm{s}}},
\end{equation}
where $c_{\rm CR}=\sqrt{\gamma_{\rm{c}}P_{\rm c}/\rho}$ is the CR acoustic speed with $\gamma_{\rm{c}}=4/3$, and $c_{\rm{s}}$ is the sound speed. Solving the equations gives the dispersion relation:
\begin{eqnarray}\label{appendixeq:disp}
    (-ikl_{\rm{c}}-m+\nu-\frac{i\gamma_{\rm{c}}}{2kL_{\rho}}m)\nonumber\bigg[\bigg(1-\frac{i}{kL_{\rho}}\bigg)\bigg(1+\frac{g}{kc_{\rm{s}}^{2}\nu^{2}}\bigg)\nu^{3}\nonumber\\
    -\nu\bigg(1-\frac{i}{k\Lg\gamma_{\rm{g}}}\bigg)-i\frac{(1-\frac{i}{kL_{\rho}})c^{2}m(\gamma_{\rm{g}}-1)}{2\Lc\gamma_{\rm{c}}k}\bigg]\nonumber\\
    -c^{2}\nu^{2}(1-\frac{i}{k\Lc\gamma_{\rm{c}}})(1+\frac{m}{\nu}(\gamma_{\rm{g}}-1))\nonumber\\
    +\frac{1}{2}m^{2}c^{2}(1-\frac{i}{kL_{\rho}})\bigg(1+\frac{3i}{2kL_{\rho}}-\frac{i}{k\Lc\gamma_{\rm{c}}}\bigg(\frac{\nu}{m}+(\gamma_{\rm{g}}-1)\bigg)\nonumber\\
    =0
\end{eqnarray}

When $kl_{\rm{c}}\gg1$ and the background is uniform (drop the $L_{\rho}$, $\Lc$ and $\Lg$ terms), without gravity $g$, Equation~\ref{appendixeq:disp} is equal to the Eq 3.18 in \citet{1994ApJ...431..689B} in 1D form.

\begin{figure}
    \centering
    \includegraphics[width=0.5\textwidth]{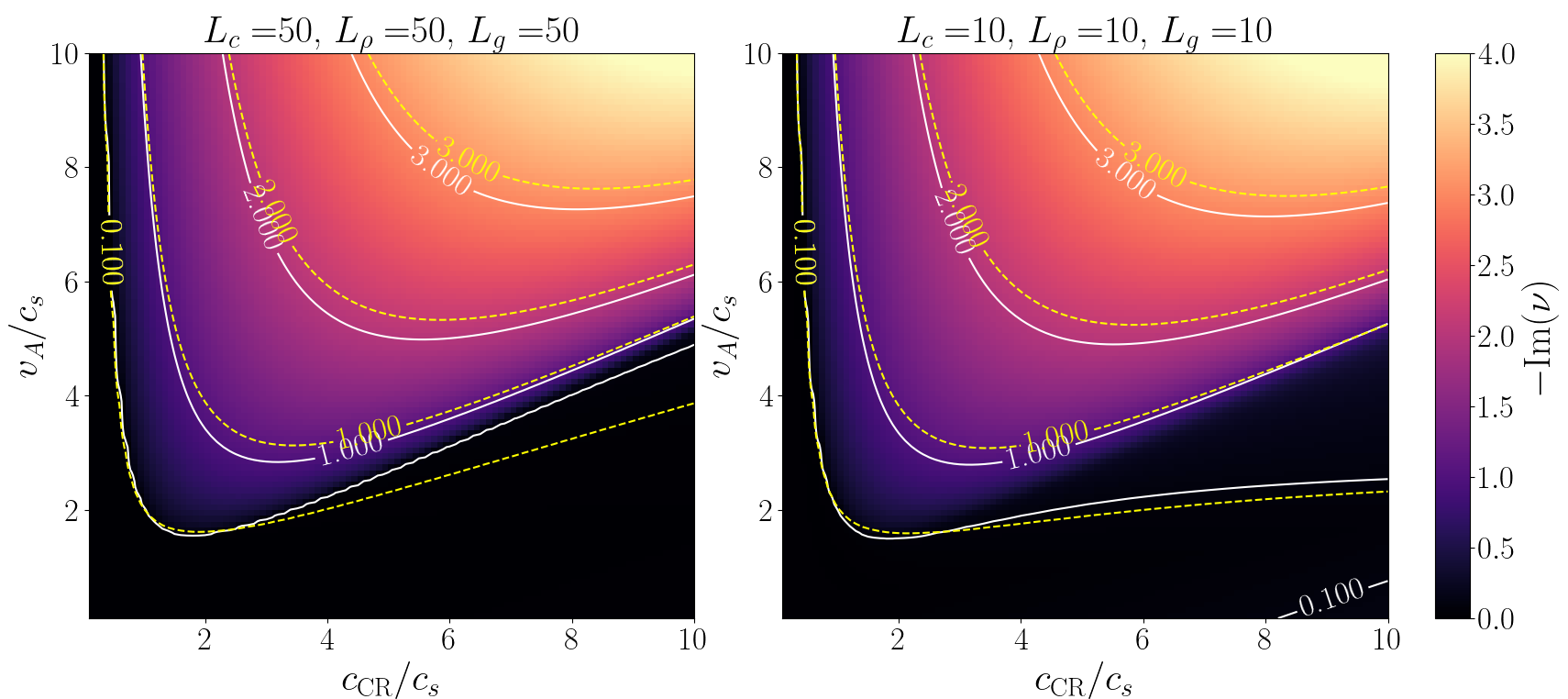}
    \caption{The re-scaled growth rate -Im($\nu$) as a function of $v_{\rm A}/c_{\rm s}$ and $c_{\rm CR}/c_{\rm s}$ for different sets of scale height $L_{\rm c}=L_{\rho}=L_{\rm g}=50$ (left) and $=10$ (right). In both plots, the gravity $g=-0.826$, the magnetic field is assumed to be uniform and static. The white solid lines represent solutions with streaming transport, the yellow dashed curves are contours represent streaming with moderate CR diffusion.}
    \label{fig:Diss_disp_comparegrad}
\end{figure}

\begin{figure}
    \centering
    \includegraphics[width=0.5\textwidth]{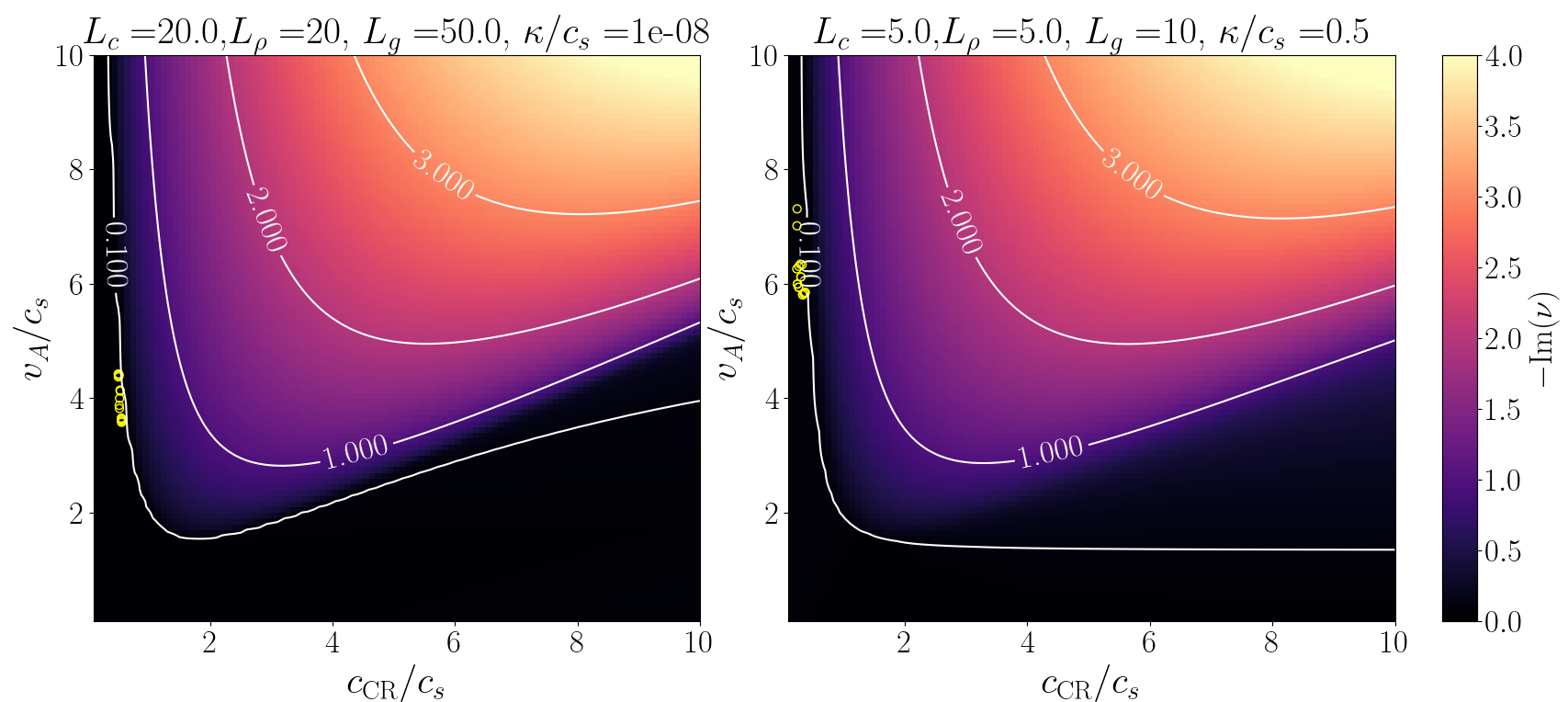}
    \caption{The re-scaled growth rate -Im($\nu$) for two of the cases shown in Figure~\ref{fig:Diss_growth}: HSE\_1F\_str (left) and HSE\_1F\_ld (right). In both plots, the white contours label the -Im($\nu$)=0.1, 1.0, 2.0 and 3.0.  The yellow circles are typical $m-c$ pair samples relevant to the instability growth, averaged over $y$ direction. }
    \label{fig:Diss_ins_weakdiffstreaming}
\end{figure}

Figure~\ref{fig:Diss_disp_comparegrad} shows the re-scaled growth rate $-\rm{Im}\nu$ for $k=1$ mode (white contour) for two sets of scale heights $\Lg=\Lc=L_{\rho}=50$ (left) and $\Lg=\Lc=L_{\rho}=10$ (right) when streaming dominates, corresponding to $l_{\rm{c}}=10^{-8}$. The yellow curves are contours for growth rate calculated with the same set of parameter, but with moderate diffusion $l_{\rm{c}}=1$. In this parameter space, adding moderate diffusion has limited impact to the growth rate.

Similarly, Figure~\ref{fig:Diss_ins_weakdiffstreaming} shows the re-scaled growth rate solution with two sets of parameters that approximate the simulation HSE\_1F\_str (left) and HSE\_1F\_ld (right). The yellow circles label $c_{\rm CR}/c_{\rm s}-v_{\rm A}/c_{\rm s}$ pairs that sampled from the diffuse region in the atmosphere at early time, where the instability emergent. The ``bottleneck'' limits CR pressure, while strong CR heating preserves gas pressure. As a result, the low $c_{\rm CR}/c_{\rm cs}$ and high $v_{\rm A}/c_{\rm s}$ constrains the growth rate to relatively low level in the simulations. 

In the left panel, we sample data from $t=6.8t_{0}$ to $t=7.2t_{0}$ with step $\Delta t=0.1t_{0}$. In each time stamp, the data points are collected at location of $x=25l_{0}, 30l_{0}, 35.0l_{0}$ averaging over the $y$ direction. In the right panel, the sampling period is $t=3.0t_{0}$ to $t=3.5t_{0}$ with step $\Delta t=0.1t_{0}$. In each time stamp, the data points are sampled at $x=20l_{0}, 25l_{0}, 30.0l_{0}$, averaging over the $y$ direction.

\end{CJK*}
\end{document}